\documentclass[aps,prb,reprint,showpacs,floatfix]{revtex4-1}
\usepackage{graphicx}  
\usepackage{dcolumn}   
\usepackage{bm}        
\usepackage{amssymb}   
\usepackage{amsmath, amsthm}
\usepackage[extra]{tipa}
\usepackage{color}
\setcitestyle{square,numbers}

\newcommand{\nag}{{\phantom{\dag}}}

\begin{document}

\title{Correlated atomic wires on substrates. II. Application to Hubbard wires}
\author{Anas Abdelwahab}
\author{Eric Jeckelmann}
\affiliation{Leibniz Universit\"{a}t Hannover, Institut f\"{u}r Theoretische Physik, Appelstr.~2, 30167 Hannover, Germany}
\author{Martin Hohenadler}
\affiliation{\mbox{Institut f\"ur Theoretische Physik und Astrophysik, Universit\"at W\"urzburg, Am Hubland, 97074 W\"urzburg, Germany}}

\date{\today}

\begin{abstract}
  In the first part of our theoretical study of correlated atomic wires on
  substrates, we introduced lattice models for a one-dimensional quantum wire
  on a three-dimensional substrate and their approximation by
  quasi-one-dimensional effective ladder models [arXiv:1704.07350]. 
  In this second part, we
  apply this approach to the case of a correlated wire with a Hubbard-type
  electron-electron repulsion deposited on an insulating substrate. The
  ground-state and spectral properties are investigated numerically using the
  density-matrix renormalization group method and quantum Monte Carlo
  simulations. As a function of the model parameters, we observe various
  phases with quasi-one-dimensional low-energy excitations localized in the
  wire, namely paramagnetic Mott insulators, Luttinger liquids, and
  spin-$1/2$ Heisenberg chains.  The validity of the effective ladder models
  is assessed by studying the convergence with the number of legs and
  comparing to the full three-dimensional model. We find that narrow ladder
  models accurately reproduce the quasi-one-dimensional excitations of the
  full three-dimensional model but predict only qualitatively whether
  excitations are localized around the wire or delocalized in the
  three-dimensional substrate.
\end{abstract}

\maketitle

\section{\label{sec:intro}Introduction}

In the first paper of this series~\cite{paper1}, we introduced a three-dimensional (3D) lattice model
for a correlated atomic wire deposited on an insulating substrate and showed how to map it onto
a two-dimensional (2D) ladder-like lattice that can be approximated by one-dimensional (1D) narrow ladder models (NLMs).
In this second paper, we apply this approach to a correlated wire represented by the 1D
Hubbard model~\cite{essler05} using density-matrix renormalization group (DMRG)~\cite{whi92,whi93,sch05,jec08a}
and quantum Monte Carlo (QMC)~\cite{rubt05,gull11} methods. 
We investigate the occurrence and properties of Luttinger liquids~\cite{sol79,giamarchi07,Schoenhammer}
and 1D Mott insulators~\cite{gebhard,giamarchi07} coupled to a substrate. 

Atomic wires on surfaces seem to be the ultimate realization of 1D electron systems~\cite{springborg07,onc08,sni10}
but the relevance of 1D physics for these materials is still controversial.
In particular, numerous experiments show that some of these materials have gapless excitation spectra with
strongly anisotropic charge dynamics. The list of good candidate materials for the realization
of (quasi-)1D conductors includes In/Si(111)~\cite{sni10}, Au/Ge(100)~\cite{blu11,blum12,naka12,park14,jong16},
Bi/InSb(100)~\cite{ohts15}, Pt/Ge(100)~\cite{yaji13,yaji16}, Pb/Si(557)~\cite{Tegenkamp2005},
and dysprosium silicide nanowires on Si(001) surfaces~\cite{Wanke2011}.
Their properties are sometimes ascribed to Luttinger liquids and sometimes to anisotropic 2D Fermi liquids.
One of the main reasons for these controversies is a poor understanding 
of the influence of the 3D substrate~\cite{springborg07,sni10,das01,das02,abd15} on 1D conductors.
Isolated 1D conductors are known to be Luttinger liquids~\cite{sol79,giamarchi07,Schoenhammer}, whereas 
the above experimental realizations raise the question of the stability of Luttinger liquids 
coupled to an environment~\cite{giamarchi07,das01,Gogolin}.

The present theoretical study sheds some light on the quasi-1D physics occurring in correlated 
atomic wires deposited on semiconducting substrates, in particular on the fate of Luttinger-liquid
and Mott-insulating phases when coupled to their environment.
In addition, it confirms that few-leg NLMs can describe---at least
qualitatively---the quasi-1D low-energy physics of the full 3D
wire-substrate system.

The paper is structured as follows. In Sec.~\ref{sec:models}, we briefly introduce
the lattice model for the wire-substrate system and its NLM approximation. 
The DMRG and QMC methods are outlined in Sec.~\ref{sec:methods}. 
Results are discussed in Sec.~\ref{sec:results} and Sec.~\ref{sec:conclusion} contains our conclusions.

\section{\label{sec:models}Models}

\subsection{3D wire-substrate model \label{sec:full_model}}

We consider a wire-substrate model that is a special case of the general model
introduced in Sec.~II of \cite{paper1}.
It consists of an interacting 1D wire on the surface of a noninteracting insulating 3D substrate.
We use a cubic lattice of size $L_x \times L_y\times L_z$
with the wire aligned in the $x$-direction and open boundary conditions in the $z$-direction.
Thus objects on the surface have a coordinate $z=0$.
We set all lattice constants and $\hbar$ equal to 1 and therefore do not
distinguish between momentum and (dimensionless) wave number. 

The system Hamiltonian can be decomposed into three terms
describing the substrate degrees of freedom, the wire degrees of freedom,
and the coupling between wire and substrate.
The 3D substrate is represented by a tight-binding Hamiltonian with a uniform nearest-neighbor hopping
$t_{\text s} > 0$ and two orbitals per site with different onsite energies $\pm \epsilon_{\text s}$
($\epsilon_{\text s} > 0$). The resulting single-particle energy spectrum has
one valence band ($\text{b}=\text{v}, \epsilon_{\text v} = - \epsilon_{\text s}$) 
and one conduction band ($\text{b}=\text{c}, \epsilon_{\text c} = + \epsilon_{\text s}$)
with the dispersion relations
\begin{equation}
\label{eq:disp}
\epsilon_{\text{b}}(\bm{k})  = \epsilon_{\text b} - 2t_{\text s} [ \cos(k_x) + \cos(k_y) + \cos(k_z) ] ,
\end{equation} 
where $k_x, k_y \in [-\pi,\pi]$ and $k_z \in [0,\pi]$. The indirect gap between the bottom of the conduction band and the top of the valence band is
$\Delta_{\text s} =  2 \epsilon_{\text s} - 12 t_{\text s} $
and the condition $\Delta_{\text s} \geq 0$ requires $\epsilon_{\text s} > 6 t_{\text s}$.

The wire is represented by the 1D Hubbard model~\cite{essler05}.
The Hubbard parameter $U\geq 0$ describes the strength of the Coulomb repulsion between
two electrons on the same site while a nearest-neighbor hopping term with
amplitude $t_{\text w}>0$ accounts for the electronic kinetic energy.  
In addition, an onsite potential $\epsilon_{\text w}=-U/2$ places the Hubbard bands
symmetrically around the middle of the substrate band gap.
For a noninteracting wire ($U=0$), we obtain the single-particle dispersion
\begin{equation}
\label{eq:wire-disp}
\epsilon_{\text w}(k_x)  =  - 2t_{\text w} \cos(k_x).
\end{equation}
The simplest coupling between the wire and the substrate consists of a 
hybridization of the electronic orbitals by a hopping term
between nearest-neighbor pairs of sites located in the wire and the
substrate, respectively. 
We use the same hybridization strength $t_{\text{ws}}$ for valence and conduction bands. 

The total Hamiltonian takes the form
\begin{eqnarray}
\label{eq:hamiltonian}
 H&=&
 -\frac{U}{2} \sum_{x,\sigma} c^{\dag}_{{\text w}x\sigma}c^{\phantom{\dag}}_{\text{w}x\sigma} 
 + U \sum_x c^{\dag}_{{\text w}x\uparrow}c^{\phantom{\dag}}_{{\text w}x\uparrow} c^{\dag}_{{\text w}x\downarrow}c^{\phantom{\dag}}_{{\text w}x\downarrow}  \nonumber \\
&& -t_{\text w}  \sum_{x,\sigma} \left ( c^{\dag}_{{\text w} x\sigma}  
c^{\phantom{\dag}}_{{\text w},x+1,\sigma} + \text{H.c.} \right ) \nonumber \\
&& + \sum_{b, \bm{r}, \sigma}  \epsilon_{\text b}  c^{\dag}_{{\text b}\bm{r}\sigma}  c^{\phantom{\dag}}_{{\text b}\bm{r}\sigma}
-t_{\text s} \sum_{\langle \bm{r} \bm{q} \rangle} \sum_{\text{b}, \sigma} \left (
c^{\dag}_{{\text b}\bm{r} \sigma}  c^{\phantom{\dag}}_{{\text b}\bm{q}\sigma} + \text{H.c.}
\right ) \nonumber \\
&& -t_{\text{ws}} \sum_{b,x,\sigma} \left ( c^{\dag}_{{\text b} \bm{r} \sigma}  
c^{\phantom{\dag}}_{{\text w} x \sigma} + \text{H.c.}  \right ) .
\end{eqnarray}
The sums over $x$ run from $1$ to $L_x$ with $\bm{r} = (x,y_0,1)$ in the last sum
($y_0$ is the $y$-coordinate of the wire), 
the sum over $\bm{r}$ runs over all substrate lattice sites,
and the sum over $\langle \bm{r} \bm{q} \rangle$ 
is over all pairs of nearest-neighbor sites in the substrate. 
The operator $c^{\dag}_{{\text b} \bm{r} \sigma}$ creates an electron with spin $\sigma$ on the site with coordinates $\bm{r} = (x,y,z)$
in the substrate orbital $\text{b}=\text{v,c}$, while $c^{\dag}_{{\text w} x\sigma}$
creates an electron with spin $\sigma$ on the wire site at $\bm{r} = (x,y_0,0)$.

\subsection{1D narrow ladder models\label{sec:nlm}}

As explained in \cite{paper1}, the full 3D wire-substrate system can be mapped 
exactly onto 
a ladder-like 2D lattice of size $L_x \times N_{\text{imp}}$
with $N_{\text{imp}}=2L_yL_z+1$ legs.
The explicit form of the full Hamiltonian is
\begin{eqnarray}
\label{eq:ladder-hamiltonian}
 H&=&
 -\frac{U}{2} \sum_{x,\sigma} g^{\dag}_{x0\sigma}g^{\phantom{\dag}}_{x0\sigma} 
 + U \sum_x g^{\dag}_{x0\uparrow}g^{\phantom{\dag}}_{x0\uparrow} g^{\dag}_{x0\downarrow}g^{\phantom{\dag}}_{x0\downarrow}  \nonumber \\
&& -t_{\text w}  \sum_{x,\sigma} \left ( g^{\dag}_{x0\sigma}  
g^{\phantom{\dag}}_{x+1,0\sigma} + \text{H.c.} \right ) \nonumber  \\
  &&-t_{\text s} \sum^{N_{\text{imp}}-1}_{n=1}\sum_{x,\sigma} \left ( g^{\dag}_{xn\sigma}g^{\phantom{\dag}}_{x+1,n\sigma} +\text{H.c.}\right) \nonumber \\
  &&-\sum^{N_{\text{imp}}-2}_{n=0} t^{\text{rung}}_{n+1}  \sum_{x,\sigma}\left( g^{\dag}_{xn\sigma}g^{\phantom{\dag}}_{x,n+1,\sigma}+\text{H.c.}\right). 
\end{eqnarray}
Here, $g^{\dag}_{xn\sigma}$ creates an electron with spin $\sigma$ at position $x$ in the $n$-th leg ($n=0,\dots,N_{\text{imp}}-1$).
The first leg ($n=0$) is identical with the wire, in particular $g^{\dag}_{x0\sigma} = c^{\dag}_{\text{w}x\sigma}$,
while legs $n=1,\dots,N_{\text{imp}}-1$ correspond to successive shells around the wire and represent the substrate. 
Hamiltonian~(\ref{eq:ladder-hamiltonian}) consists of the original Hubbard Hamiltonian for the wire, 
an intra-leg hopping $t_{\text s}$ in every substrate leg, and a nearest-neighbor rung hopping $t^{\text{rung}}_n$ between substrate legs $n-1$ and $n$.
The first two rung hoppings are  $t^{\text{rung}}_{1}=\sqrt{2}t_{\text{ws}}$ and
$t^{\text{rung}}_{2}=\sqrt{3t^2_{\text s}+\epsilon_{\text s}^2}$.
For larger $n$, $t^{\text{rung}}_{n+1}$ can be computed numerically using the Lanczos algorithm as described
in Sec.~III of \cite{paper1}.
The relation between $c^{\dag}_{{\text b} \bm{r} \sigma}$
and $g^{\dag}_{xn\sigma}$ is also explained there.

The mapping of the 3D wire-substrate model to the 2D ladder-like system is
exact but does not yet simplify the
problem.
Intuitively, however, 1D physics (such as Luttinger liquid behavior) should occur in the wire
or in a region of the substrate around the wire. Thus
only legs that are close to the wire should be essential for a qualitative description of 
the 1D low-energy properties.
Therefore, we approximate the 3D wire-substrate model by effective NLMs that
are obtained by taking only the $N_{\text{leg}} \ll  N_{\text{imp}}$ legs
closest to the wire into account.
The investigation of a noninteracting wire in \cite{paper1} established that
an NLM must include an odd number of legs $N_{\text{leg}} \geq 3$ to describe
a wire on an {\em insulating} substrate.

\subsection{Parameters}

For insulating substrates, we can find model parameters such that
the low-energy excitations of the noninteracting wire lie in the substrate band gap.
These excitations are then localized on or around the wire and thus form
 a 1D electronic subsystem of the full 3D wire-substrate system.
 In \cite{paper1} we showed that this scenario is achieved
 at half-filling and close to half-filling with a wire hopping $t_{\text{w}}=3$
 and the substrate parameters $t_{\text{s}}=1$ and 
$\epsilon_{\text{s}}=7$.
The latter correspond to an indirect gap $\Delta_{\text s} =  2 \epsilon_{\text s} - 12 t_{\text s}=2$ 
and a direct gap $\Delta(k_x) = 2 \epsilon_{\text s} - 8 t_{\text s} = 6$ for a fixed wave number $k_x$
in the single-particle excitation spectrum. 

The effective substrate band gap $\Delta_{\text{s}}(N_{\text{leg}})$ is larger in the NLM but converges to $\Delta_{\text s}$ for 
$N_{\text{leg}} \rightarrow \infty$. For instance, for the three-leg NLM
at vanishing wire-substrate coupling $t_{\text{ws}}=0$, the substrate is represented by a noninteracting two-leg ladder with 
single-particle energies
\begin{equation}
\epsilon(k_x)  =  \pm t^{\text{rung}}_{2}  - 2t_{\text s} \cos(k_x).
\end{equation}
Thus $\Delta_{\text{s}}(N_{\text{leg}}=3) =  2t^{\text{rung}}_{2}  - 4t_{\text s}  \approx 10.4$ 
is five times larger than the true gap $\Delta_{\text{s}} = 2$.
We use the above parameters throughout this work
and focus on the model properties as a function of  the hybridization
between wire and substrate $t_{\text{ws}}$ and the strength of the electron-electron interaction $U$.

At half-filling, the 3D wire-substrate model contains $N_p=N_{\text{imp}}L_x$
electrons, whereas the NLM contains  $N_p=N_{\text{leg}}L_x$. 
We focus on half-filled systems and on systems doped away from half-filling
by a finite wire doping $y_{\text w} \in (-1,1)$
($N_p = N_{\text{imp}}L_x + y_{\text w}L_x$ for the 3D wire-substrate model,
or $N_p=N_{\text{leg}}L_x + y_{\text w}L_x$ for the NLM).
Such a finite wire doping corresponds to a negligible bulk doping of the substrate in the thermodynamic limit
$N_{\text{imp}} \gg 1$ but is relevant for quasi-1D conductors embedded in an
insulating 3D bulk system, e.g., metallic wires on semiconducting substrates.

\section{Methods \label{sec:methods}}

\subsection{DMRG \label{sec:DMRG}}

The DMRG is a powerful method for quasi-1D correlated quantum systems with short-range interactions~\cite{whi92,whi93,sch05,jec08a}.
It can be used to study relatively wide ladder geometries~\cite{hage05}
or coupled chains~\cite{mouk04,mouk10}.
For such systems, however, it is limited by an exponential increase  
of CPU time and required memory as a function of the lattice width. 
Therefore, our DMRG study is necessarily restricted to correlated NLMs
with small numbers of legs $N_{\text{leg}}$. 
Nevertheless, we found that in general the computational effort required for the NLM 
increases much more slowly with the number of legs than for a similar homogeneous ladder system.
Fundamentally, the exponential increase of the computational cost 
is due to the rapid increase of entanglement with the ladder width.
This entanglement is essentially determined by the number of gapless excitation modes
in the system
(e.g., the number of bands crossing the Fermi energy in a noninteracting system).
In a homogeneous ladder model, this number is typically proportional to the ladder width.
In the NLM for an insulating substrate, however, this number remains small when $N_{\text{leg}}$
increases because most excitation modes represent gapped transitions between valence and conduction bands.
This results in a slower increase of the computational cost with system width. 

We used the finite-system DMRG algorithm on lattices with up 
to $L_x=208$ rungs for three-leg ladders and up to $L_x=128$
for wider ladders with up to $N_{\text{leg}}=11$ legs.
The ladder length $L_x$ was always taken to be an even number
and open boundary conditions were used in the $x$-direction.  
Up to $m=2024$ density-matrix eigenstates
were kept in our DMRG calculations, yielding discarded weights
smaller than $10^{-6}$. 
We systematically investigated truncation errors by keeping variable numbers
of density-matrix eigenstates and extrapolating ground-state energies to the limit
of vanishing discarded weights~\cite{bonc00}.
The resulting error estimates are smaller than the symbols in our figures.

Using the DMRG, we calculated the charge gap
\begin{eqnarray}
\label{eq:Ec}
E_{\text{\text c}} & = & \frac{1}{2} \left [ E_{0}(M_{\uparrow}+1,M_{\downarrow}+1)
+E_{0}(M_{\uparrow}-1,M_{\downarrow}-1) \right . 
\nonumber \\ 
&& \hspace*{1em}\left . -2E_{0}(M_{\uparrow},M_{\downarrow}) \right ] \,,
\end{eqnarray}
the spin gap
\begin{equation}
\label{eq:Es}
E_{\text{s}} = E_{0}(M_{\uparrow}+1,M_{\downarrow}-1) - E_{0}(M_{\uparrow},M_{\downarrow})\,, 
\end{equation}
and the single-particle gap
\begin{eqnarray}
\label{eq:Ep}
E_{\text{p}} & = & E_{0}(M_{\uparrow}+1,M_{\downarrow})+E_{0}(M_{\uparrow}-1,M_{\downarrow})
\nonumber \\ 
&& \hspace*{1em} -2E_{0}(M_{\uparrow},M_{\downarrow}) \,,
\end{eqnarray}
where $E_{0}(M_{\uparrow},M_{\downarrow})$ denotes the ground-state
energy for $M_{\sigma}$ electrons of spin $\sigma$. 
These gaps are visible in the dynamic charge structure factor,
the dynamic spin structure factor, and the single-particle spectral functions
calculated with the QMC method discussed below.

Additional information can be inferred from the distribution of charges and spins on the different legs.
The total charge on leg $n$ is defined as
\begin{equation}
\label{eq:chargedens}
C(n)=\left \langle\psi_{\text{GS}} \left \lvert \sum_{x,\sigma}g^{\dag}_{xn\sigma}g^{\phantom{\dag}}_{xn\sigma}  
\right \rvert\psi_{\text{GS}} \right \rangle\,,
\end{equation}
while the total spin-$z$ density is defined by 
\begin{equation}
\label{eq:spindens}
S(n)=\left \langle\psi_{\text{GS}} \left \lvert \sum_{x,\sigma} \sigma g^{\dag}_{xn\sigma}g^{\phantom{\dag}}_{xn\sigma}  
\right \rvert\psi_{\text{GS}} \right \rangle\,.
\end{equation}
Here, $\left \rvert\psi_{\text{GS}} \right \rangle$ is the ground state for
$M_\sigma$ electrons of spin $\sigma$. Additionally, variations 
$\Delta C(n)$ and $\Delta S(n)$ of these quantities
for $M_{\sigma}\pm 1$ indicate whether the lowest charge, spin and single-particle excitations
(defined by the above gaps)
are mostly localized on the wire or distributed in the substrate.

The actual excess density on the wire is
\begin{equation}
\label{eq:eff_doping}
y_{\text{eff}} = \frac{C(0)}{L_x}-1
\end{equation}
while the wire doping corresponds to 
\begin{equation}
\label{eq:wire_doping}
y_{\text w} = \sum_{n=0}^{N_{\text{leg}}-1} \left [ \frac{C(n)}{L_x}-1 \right ] .
\end{equation}
If all added electrons (or added holes) are localized in the wire then
$y_{\text{eff}} = y_{\text w}$. However, in general, $\vert y_{\text{eff}} \vert <  \vert y_{\text w} \vert$ because the doped particles 
have a finite probability to be in the substrate. We will show below that it is possible that
they become completely delocalized in the whole substrate so that $\vert y_{\text w} \vert \gg \vert y_{\text{eff}} \vert \approx 0$.

\begin{figure}[t]
\includegraphics[width=0.4\textwidth]{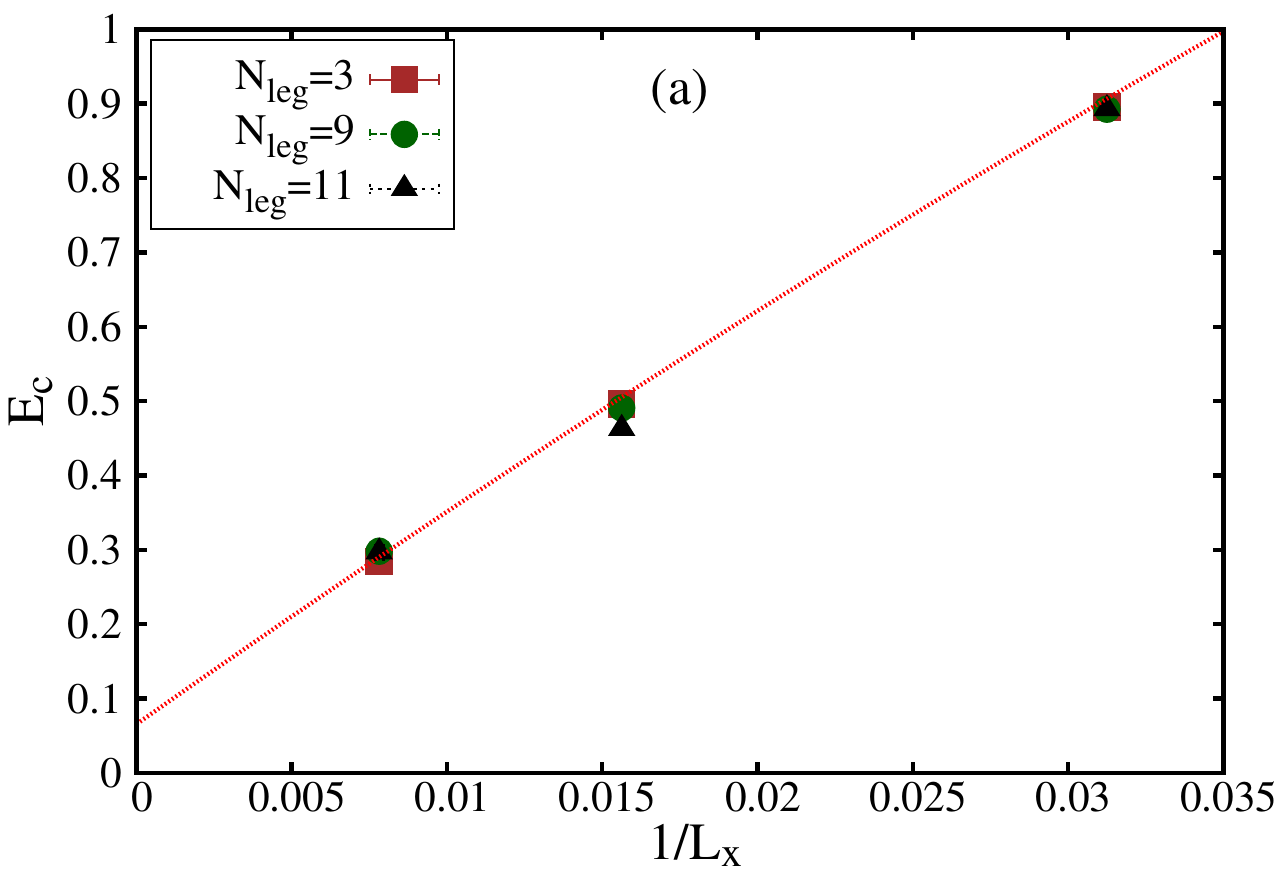}
\includegraphics[width=0.4\textwidth]{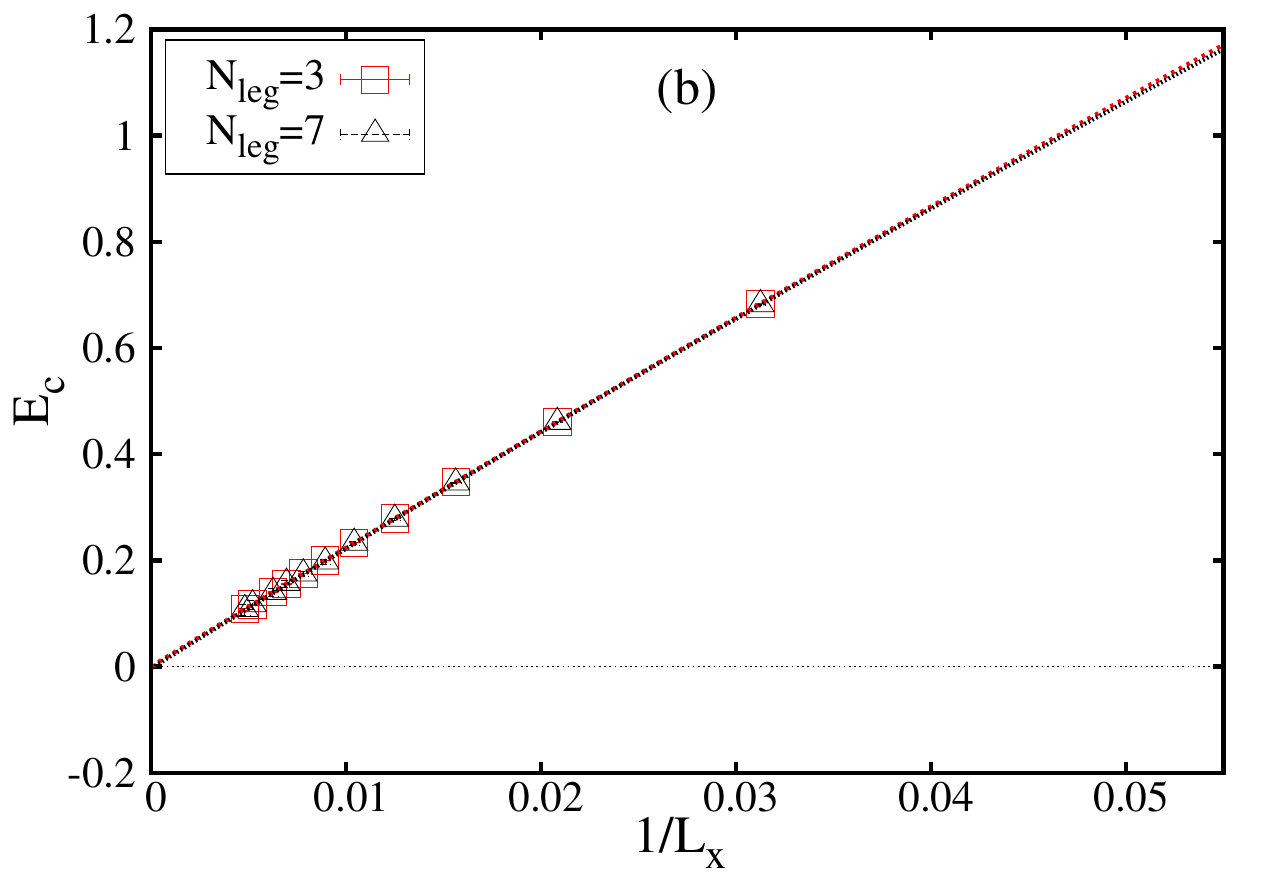}
\caption{\label{figgap_Ec_all} (Color online)
DMRG results for the charge gap [Eq.~(\ref{eq:Ec})] of the NLM 
(a) at half-filling and (b) away from half-filling ($y_{\text w}=12.5$\%)
as a function of the inverse ladder length $1/L_x$ for different numbers of legs $N_{\text{leg}}$.
The Hubbard interaction was $U=4$.
The line in (a) shows the finite-size gap of the half-filled 1D Hubbard model
with open boundary conditions and a hopping $t_{\text{w}}=3$. The lines in (b) 
correspond to quadratic fits in $1/L_x$.
}
\end{figure}

To analyze finite-size corrections in the correlated NLM, we
calculated these gaps for ladders of various 
lengths $L_x$ and widths $N_{\text{leg}}$.
As a first example, Fig.~\ref{figgap_Ec_all}(a) shows the charge gap 
$E_{\text{\text c}}$ at half-filling for $U=4$ as a 
function of $1/L_x$ for different $N_{\text{leg}}$. 
It decreases as a function of $1/L_x$
for a fixed $N_\text{leg}$.
In 1D Mott insulators, charge and single-particle gaps
decrease
toward finite values in the limit $L_x\rightarrow\infty$.
We used a second-order polynomial fit in $1/L_x$ to 
extrapolate these gaps whenever necessary and possible. 
In Fig.~\ref{figgap_Ec_all}(a) the extrapolated charge gaps
are finite and almost equal for all $N_{\text{leg}}$ for the parameters chosen. 
For other model parameters, we find that the gap can strongly depend
on the number of legs and that extrapolations for increasing $N_{\text{leg}}$ at finite system length $L_x$
are also unsatisfactory. 
This complex finite-size scaling is related to the large variation of the effective
substrate band gap $\Delta_{\text{s}}(N_{\text{leg}})$ with $N_{\text{leg}}$
discussed before for the noninteracting NLM.

As a second example, Fig.~\ref{figgap_Ec_all}(b) 
shows the charge gap away from half-filling ($y_{\text w}=12.5$\%) for $U=4$
for two different values of $N_{\text{leg}}$.
The charge gap vanishes with $1/L_x$ for a fixed number of legs.
According to conformal field theory, the
finite-size gaps of gapless excitations in 1D electron systems
vanish linearly with the inverse of the system length~\cite{giamarchi07},
\begin{equation}
\label{eq:velocities}
E_\alpha = \frac{\pi v_\alpha}{L_x}
\end{equation}
for $L_x \gg 1$, where $\alpha=\text{c}$, s, or p and $v_\alpha$ is the velocity of the corresponding excitation. 
We can hence calculate the velocities of charge, spin and single-electron excitations from
the line slopes in the finite-size-scaling analysis. For the
noninteracting wire without a substrate these velocities are
equal to the Fermi velocity $v_{\text{F}} = 2t_{\text{w}} \sin(k_{\text{F}})$, 
with $v_{\text{F}} = 2t_{\text{w}} = 6$ at half-filling and 
$v_{\text{F}} \approx 1.96 t_{\text{w}} \approx 5.88$ 
at $12.5\%$ doping.
We find that the $v_\alpha$ do not change significantly with the number of 
legs for $N_{\text{leg}} \geq 3$, as illustrated in
Fig.~\ref{figgap_Ec_all}(b) where $v_\text{c}/v_{\text{F}} \approx 1.2$.
Since our DMRG results for gapless excitation modes are limited to a few values of $N_{\text{leg}}$,
we can in principle not rule out more significant finite-size corrections for $N_{\text{leg}}>7$.

Ideally, the finite-size gaps of the NLM should be extrapolated to the thermodynamic limit
using a fixed ratio $N_{\text{leg}}/L_x$. However, this is not possible with
the DMRG because we cannot simulate enough different values of $N_{\text{leg}}$
for fixed $N_{\text{leg}}/L_x$.
Therefore, the few-leg correlated NLMs accessible to the DMRG are not large
enough to accurately investigate the full 2D ladder representation~(\ref{eq:ladder-hamiltonian}) of the wire-substrate system. However, they can yield
a useful approximation,  as illustrated by our observation that 
the essential properties (e.g., gapped vs. gapless excitations, or excitations on the wires vs. in the substrate)
do not change significantly with $N_{\text{leg}}$ for $N_{\text{leg}}\geq 3$ 
unless a phase boundary is crossed.

To achieve larger ladder sizes, one could use other DMRG methods and other representations of the NLM 
that are more appropriate for specific problems. 
For instance, the two-step DMRG~\cite{mouk04,mouk10} allows one to investigate
systems of weakly coupled chains more efficiently.
The DMRG can also be used in momentum space~\cite{xian96,nish02,ehle15}, where it yields more accurate results 
for momentum-resolved observables for weak electron-electron interactions. As the NLM is not translationally
invariant in the rung direction, however, this approach is not directly applicable.
Nevertheless, a clear advantage of the momentum representation of the NLM is that the $yz$-slices of the substrate 
are decoupled, see Sec.~III of \cite{paper1}.
It is sufficient to use the momentum representation in the wire direction ($x$-direction)
to achieve this decoupling. Thus one could also envision using a mixed representation ($k_x,y,z$), 
i.e., momentum space in the wire direction and real space in the $y$- and $z$-directions, or 
($k_x,n$), i.e., momentum space in the wire direction and Lanczos basis for the other two directions.
DMRG variants that combine momentum and real space have been developed recently 
to take advantage of such alternative representations~\cite{motr16,ehle17}.
The mixed representation $(k_x,n)$ is expected to be the best starting point for field-theoretical 
approaches~\cite{sol79,giamarchi07,Gogolin,Tsvelik}.

Alternatively, it is possible to consider each $yz$-slice of the substrate (or, equivalently, each rung of the NLM) 
as a single site with a large number of states and apply DMRG methods developed to treat such big sites~\cite{zhan98,burs99,jeck07}.
This approach may lead to much smaller effective representations of the substrate degrees of freedom
because the latter seem to be more weakly entangled than the rungs of homogeneous ladder systems.

\subsection{QMC \label{sec:QMC}}

The continuous-time interaction-expansion (CT-INT) QMC method \cite{rubt05} is particularly
useful to study both NLMs and the full 3D wire-substrate model. For this purpose, the method is formulated in terms
of the fermionic coherent-state path integral with an action $S = S_0 +
S_1$. Here, $S_0$ is quadratic and has the form
\begin{equation}\nonumber \label{eq:action}
  S_0 = -\sum_{ij\sigma} \iint_{0}^{\beta} d\tau d\tau'\,
  {c}^{\dagger}_{i\sigma}(\tau) G^{-1}_{0,\sigma} (i-j,\tau-\tau')
  {c}^\nag_{j\sigma}(\tau')\,,
\end{equation}
with the free Green function ${G}_{0,\sigma}$ describing the hopping between
sites $i$ and $j$ of the wire via all possible paths (direct or via the substrate).
The Hubbard interaction in the wire is contained in 
\begin{equation}\nonumber \label{eq:action2}
  S_1 = 
  {U} \sum_{i} \int_{0}^{\beta} d\tau
  \left[{c}^{\dagger}_{{i}\uparrow}(\tau) {c}^\nag_{{i}\uparrow}(\tau) - \frac{1}{2} \right]
  \left[{c}^{\dagger}_{{i}\downarrow}(\tau) {c}^\nag_{{i}\downarrow}(\tau) - \frac{1}{2} \right]
  \,.
\end{equation}
The key idea of the method is a Dyson-expansion of the partition function $Z =
\text{tr}\, e^{-S}$ in powers of $S_1$, which can be summed exactly by
stochastic sampling of interaction vertices \cite{rubt05}. The algorithmic details have
been discussed in detail before~\cite{gull11}. For the present problem, it is essential to
understand that interactions are restricted to the wire, whereas substrate
sites are noninteracting both in the NLM and the full 3D problem. As in
previous work on edge states of topological insulators \cite{HoLaAs10}, the
numerical effort scales as $n^3$, where $n\sim
U \beta L_x$ is the average expansion order and  depends only on the number
of correlated sites. It is hence the same as for the 1D Hubbard model. The cubic scaling with $n$ makes CT-INT most
useful for weak to intermediate couplings. However, because the noninteracting substrate sites are
integrated out, NLMs and full 3D models with the same $L_x$ require the same
computer time so that detailed comparisons between these different models are possible.

Here, we used a grand-canonical variant of CT-INT with inverse temperature
$\beta$. A chemical potential $\mu=0$ corresponds to half-filling,
whereas $\mu>0$ gives electron-doped systems with $y_{\text w}>0$. The total
number of doped electrons was adjusted to $L_x/8=5.25$ (or $y_{\text
  w}\approx 12.5\%$, as in the
DMRG results) by tuning $\mu$. The thermodynamic average of the particle density
was calculated exactly for the wire as well as the substrate legs of the
three-leg NLM [cf. Eqs.~(\ref{eq:chargedens}) and~(\ref{eq:wire_doping})]. For
the 3D wire-substrate model, substrate averages over all sites are not
feasible because $G_{0}$ has to be stored for all sites and
imaginary times. Therefore, substrate properties were obtained by averaging
over the chains at minimal ($y=1$, $z=1$) and maximal distance ($y=L_y/2$, $z=L_z$)
from the wire.

To complement the DMRG results, we calculated spectral properties of the NLM
and the 3D wire-substrate system. Specifically, we considered the momentum-
and energy-resolved single-particle spectral functions, as well as the dynamic charge
and spin structure factors. These quantities 
can be measured in experiments such as angle-resolved  photoemission spectroscopy,
electron-energy-loss spectroscopy, and inelastic neutron scattering, respectively.
In principle, dynamic quantities are also accessible with the DMRG
method~\cite{ben04,ben07,jec08} but at a high computational cost and with
the additional complication of using pseudo-wave numbers and open boundary
conditions (see Refs.~\cite{noce16a,noce16b} for recent works).

We considered the single-particle spectral functions defined in
\cite{paper1}, namely the wire spectral function $A_{\text{w}}(\omega,k_x)$, 
the ``substrate'' spectral function $A_{\text{s}}(\omega,k_x)$ of the three-leg NLM,
and the substrate spectral function $A_{\text s}(\omega,k_x)$ corresponding
to the average of $A_{\text s}(\omega,k_x,y=y_0,z=1)$ and $A_{\text s}(\omega,k_x,y=y_0+L_y/2,z=L_z)$.

The dynamic charge ($\alpha=\rho$) and spin ($\alpha=\sigma$) structure
factors of the wire are defined as
\begin{align}\label{eq:akw}\nonumber
  S_{\alpha}(\omega,k_x) 
  &=
  \frac{1}{Z}\sum_{ij}
  {|\langle {i}| \hat{S}_{\alpha}(k_x) |{j}\rangle|}^2 (e^{-\beta E_i}+e^{-\beta E_j})
  \\
  &\hspace*{4em}\times
  \delta(E_j-E_i-\omega)
  \,.
\end{align}
with 
\begin{align}\nonumber
\hat{S}_{\rho}(k_x)  &= \frac{1}{\sqrt{L_x}} \sum_{x} e^{ik_xx} \sum_{\sigma}  c^{\dag}_{{\text w} x \sigma} c^{\phantom{\dag}}_{{\text w} x \sigma}\,,  
\\
\hat{S}_{\sigma}(k_x) &= \frac{1}{\sqrt{L_x}} \sum_{x} e^{ik_xx} \sum_{\sigma} \sigma c^{\dag}_{{\text w} x \sigma} c^{\phantom{\dag}}_{{\text w} x \sigma} \,.
\end{align}
Here, $|{i}\rangle$ is an eigenstate with energy $E_i$.
The above spectral functions were determined from the QMC results
for the corresponding single-particle, density-density and spin-spin
imaginary-time Green functions with the help of the stochastic maximum
entropy method~\cite{bea04}.

\section{Results \label{sec:results}}

\subsection{Insulating wire}

For half-filling and $t_{\text{ws}}=0$, the wire is an exactly half-filled Hubbard chain decoupled from the substrate.
The ground state of this 1D model for repulsive interactions is a paramagnetic Mott insulator~\cite{essler05}.
Therefore, we know that the 3D wire-substrate model and the NLM are Mott insulators if $U>0$ and $t_{\text{ws}}=0$.

\begin{figure}[t]
\includegraphics[width=0.4\textwidth]{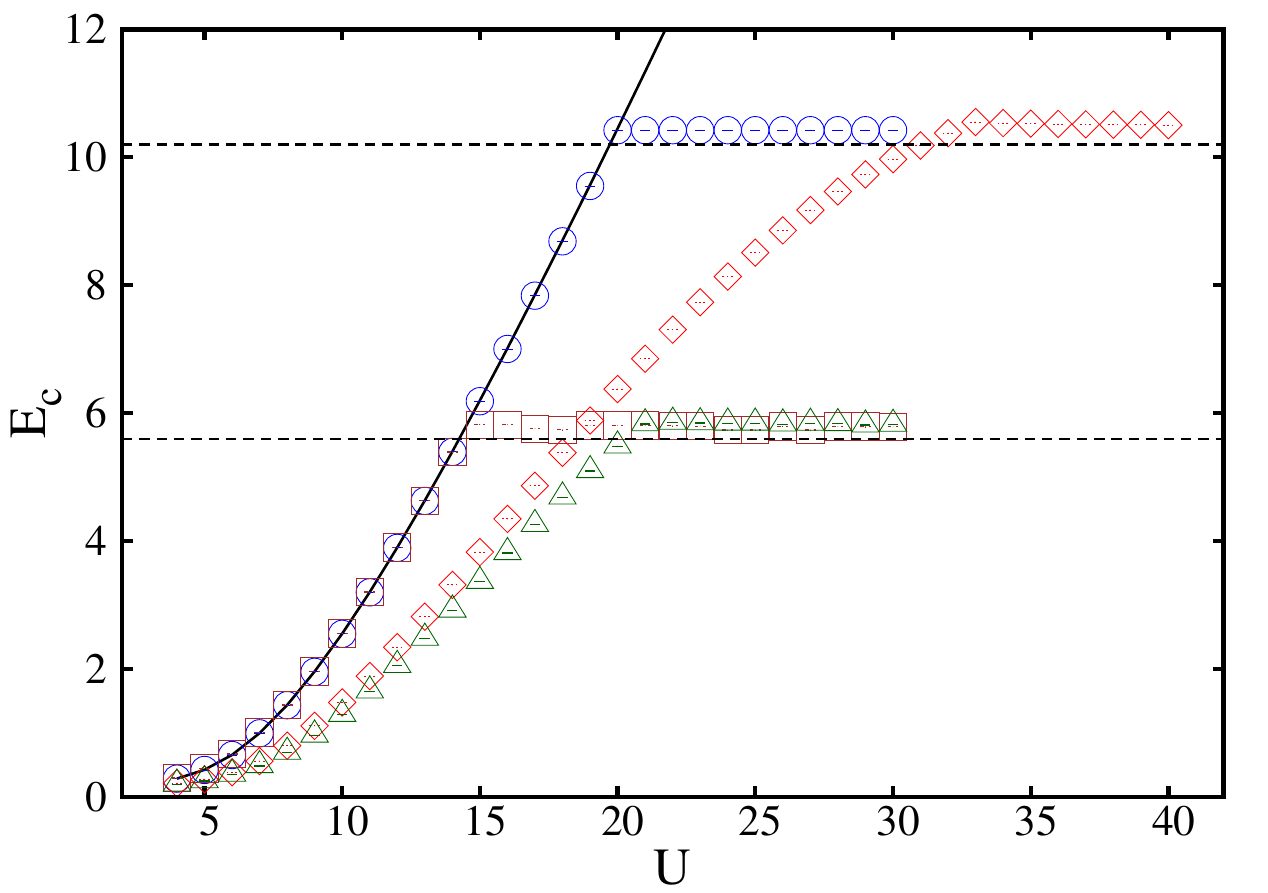}
\caption{\label{fig:Ec_vs_U} (Color online)
DMRG results for the charge gap [Eq.~(\ref{eq:Ec})] at half-filling as a function of the Hubbard interaction $U$
for a three-leg NLM with $t_{\text{ws}}=0.1$ (circles)
and $t_{\text{ws}}=2$ (diamonds), as well as for a seven-leg NLM with $t_{\text{ws}}=0.1$ (squares)
and $t_{\text{ws}}=2$ (triangles).
The solid line indicates the Mott gap of the 1D Hubbard chain with a hopping $t_{\text{w}}=3$.
The horizontal dashed lines indicate the effective substrate band gaps 
$\Delta_{\text{s}}(N_{\text{leg}}=3)\approx 10$ and $\Delta_{\text{s}}(N_{\text{leg}}=7)\approx 5.5$
of the noninteracting NLMs.
}
\end{figure}

Figure~\ref{fig:Ec_vs_U} shows the charge gap $E_{\text c}$ 
of half-filled three-leg and seven-leg correlated NLMs as
a function of the interaction $U$ for $L_x=128$ and $t_{\text{ws}}>0$. 
Finite-size effects are considerable for small charge gaps (i.e., small $U$) 
but our finite-size analysis show that $E_{\text c}$ is finite in the thermodynamic limit
at least for $U\geq 4$, see Fig.~\ref{figgap_Ec_all}. For weaker interactions, we could not 
determine if the charge gap remains finite for $L_x\rightarrow \infty$. At 
stronger interactions, finite-size effects are smaller than the symbol size
in Fig.~\ref{fig:Ec_vs_U}. For weak hybridizations $0<t_{\text{ws}}\alt 0.5$,
$E_{\text c}$ increases with $U$ almost exactly
as the Mott gap of a Hubbard chain~\cite{essler05} up to
$U_{{c}}\approx20$, before saturating abruptly at
a value close to the effective substrate band gap for the three-leg NLM 
[$\Delta_{\text{s}}(N_{\text{leg}}=3) \approx 10$].
For stronger wire-substrate hybridization, such as $t_{\text{ws}}=2$ in Fig.~\ref{fig:Ec_vs_U},
the charge gap becomes smaller than the Mott gap of the Hubbard chain but its
dependence on $U$ remains qualitatively the same, with saturation occurring
at a slightly larger gap value $E_{\text c}$ and thus at a larger $U_{{c}}$. 
A finite charge gap in the thermodynamic limit and a saturation effect can be
observed for hybridizations up to $t_{\text{ws}}=4$. 
Finally, Fig.~\ref{fig:Ec_vs_U} shows that the behavior of the gap is qualitatively similar for the three-leg and the seven-leg NLM
but that the critical coupling $U_{{c}}$ decreases with increasing $N_{\text{leg}}$.
The single-particle gap behaves essentially like the charge gap. 

\begin{figure}
\includegraphics[width=0.4\textwidth]{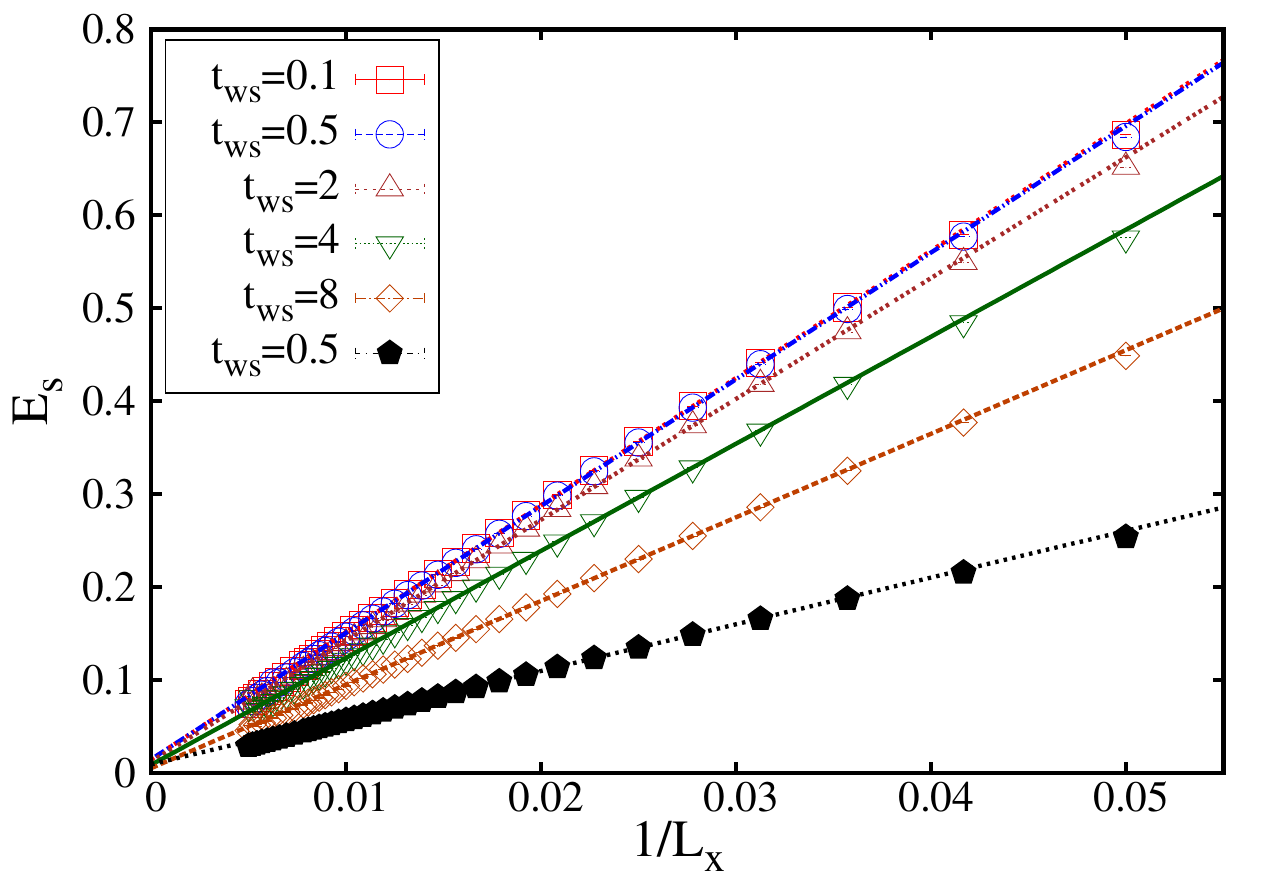}
\caption{\label{fig:Es_vs_L} (Color online) DMRG results for the 
spin gap [Eq.~(\ref{eq:Es})] of the half-filled three-leg NLM as a function
of the inverse ladder length for $U=4$ (open symbols) and $U=24$ (filled
symbols) and various hybridization strengths $t_{\text{ws}}$.
}
\end{figure}

Figure~\ref{fig:Es_vs_L} shows that the spin gap vanishes linearly with $1/L_x$ in the three-leg NLM, as expected for a 
1D Mott-Hubbard insulator or a spin-$1/2$ Heisenberg chain. 
This scaling is observed both above and below the charge gap saturation value $U_c$.
The slopes (i.e., spin velocities) decrease with increasing $t_{\text{ws}}$,
suggesting that the effective exchange coupling between spin degrees of freedom becomes weaker.
This behavior remains qualitatively similar for larger $N_{\text{leg}}$.

\begin{figure}[b]
\includegraphics[width=0.4\textwidth]{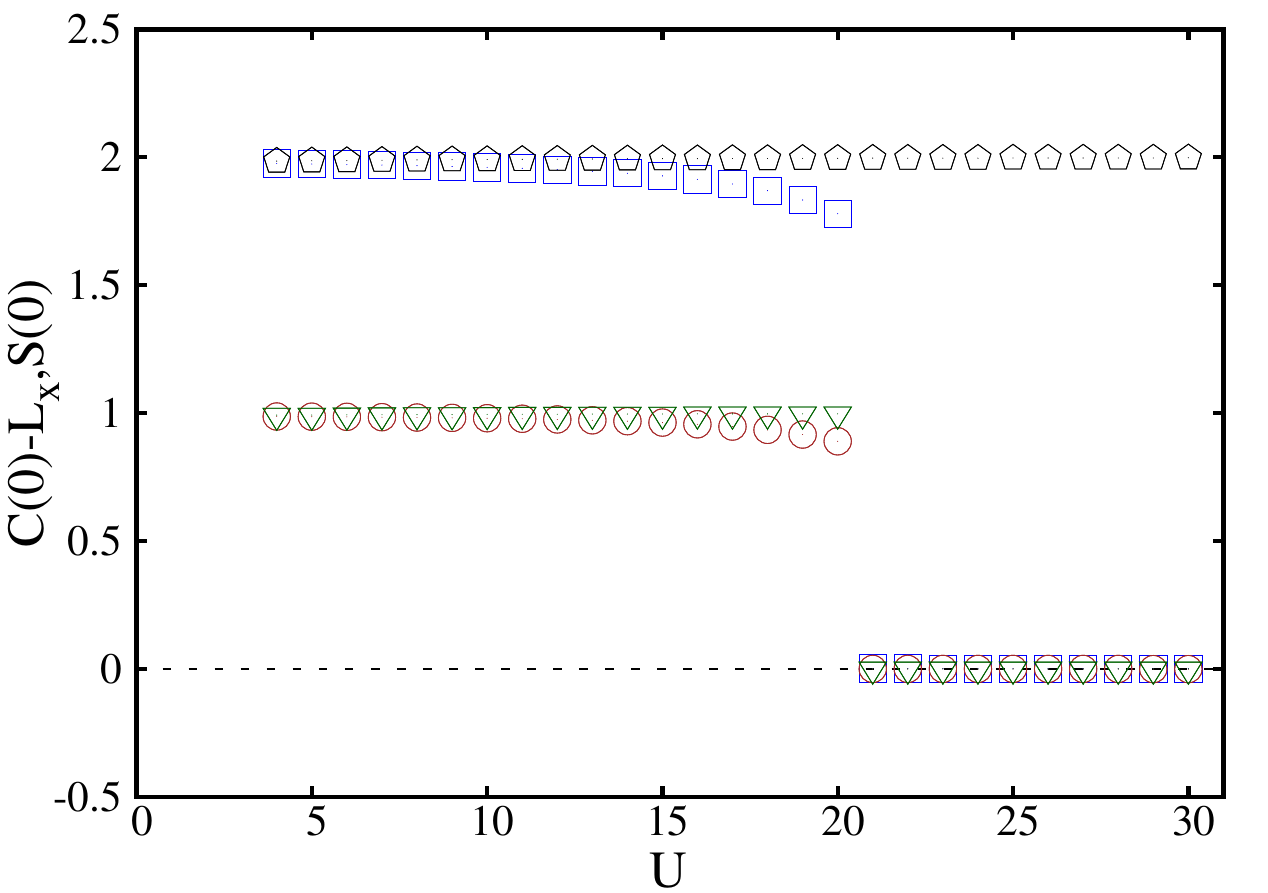}
\caption{\label{fig:densitiesHF} (Color online)
DMRG results for the variation of the total charge $\Delta C(0) = C(0)-L_x$
and spin $\Delta S(0)=S(0)$ on the wire leg of the half-filled three-leg NLM
as a function of the Hubbard interaction $U$. Here, $t_{\text{ws}}=0.5$.
The different symbols correspond to the lowest charge (squares), spin (pentagons), and
single-particle (circles and triangles, respectively) excitations.
}
\end{figure}

The charge and spin distributions~(\ref{eq:chargedens}) and~(\ref{eq:spindens})
of the half-filled ground state are featureless. 
Figure~\ref{fig:densitiesHF} shows 
that the variations of these quantities for
the lowest excitations provide much more information.
The variations of $C(n)$
for one or two added electrons reveal that the lowest charge excitations are mostly situated on the
wire leg for $U < U_{{c}}$ but on the noninteracting substrate legs for $U > U_{{c}}$.
The variations of $S(n)$ for a triplet excitation shows that the lowest spin excitations are localized on the wire leg
for any $U \geq 4$. In contrast, for a single-particle excitation (i.e., one added electron), the excess spin goes on the wire leg for
$U < U_{{c}}$ but on the substrate legs for $U > U_{{c}}$.
These uneven distributions are more pronounced for weaker hybridizations $t_{\text{ws}}$.
We have verified that they remain qualitatively similar for
larger numbers of legs up to $N_{\text{leg}}=7$. 

\begin{figure}
\includegraphics[width=0.5\textwidth]{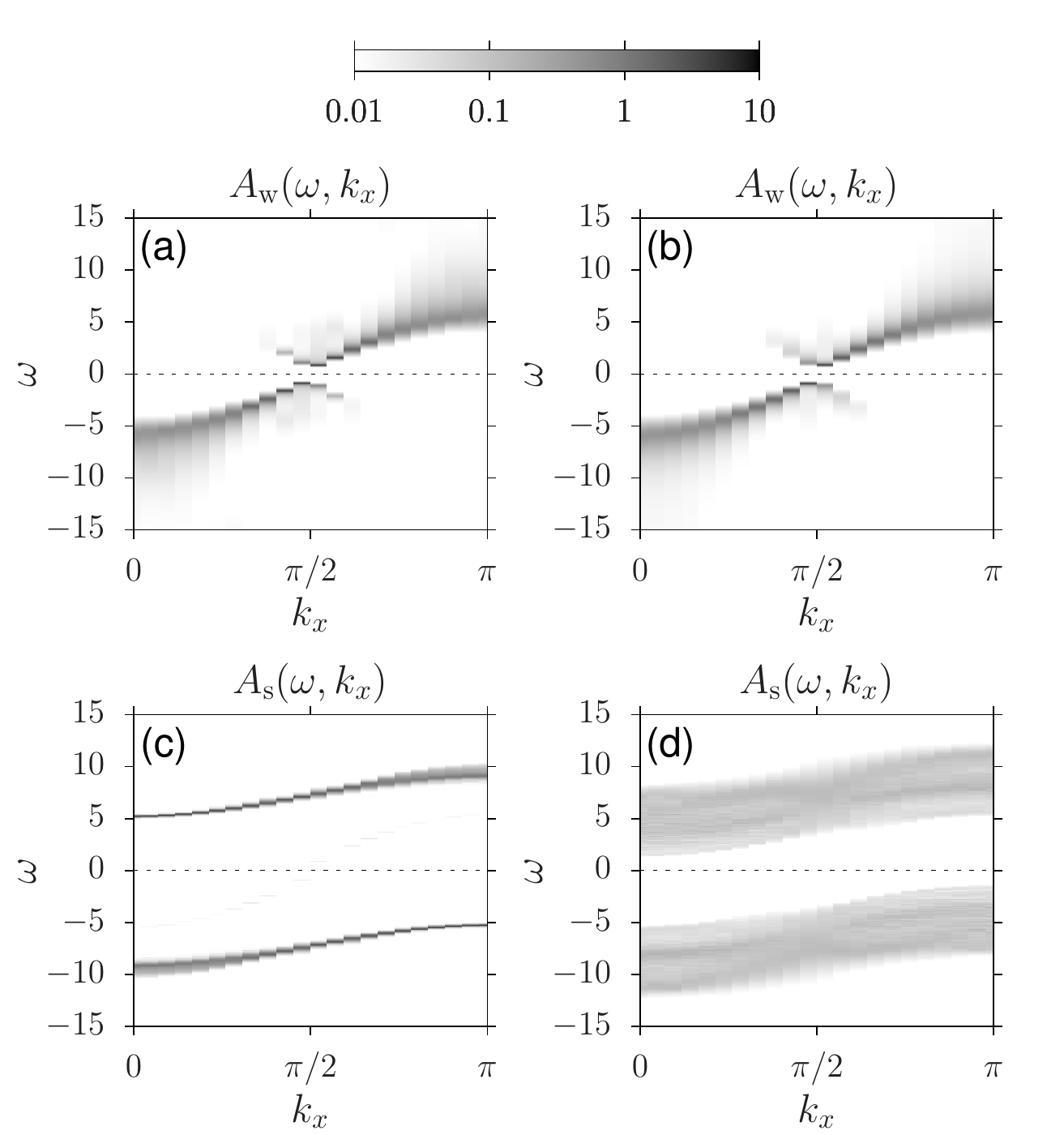}
\caption{\label{fig:qmc_U8}
CT-INT results for the spectral functions $A_{\text{w}}(\omega,k_x)$ [(a),(b)] and 
$A_{\text{s}}(\omega,k_x)$ [(c),(d)] for $U=8$, $t_{\text{ws}}=0.5$, $\beta=15$, and $L_x=42$. 
The chemical potential was $\mu=0$, corresponding to half-filling. Panels (a)
and (c) show results for the three-leg NLM, panels (b) and (d) for the 3D wire-substrate model ($L_y=42$, $L_z=10$). 
}
\end{figure}

The CT-INT single-particle spectral functions are shown in Fig.~\ref{fig:qmc_U8} for
the three-leg NLM and the 3D wire-substrate model with $t_{\text{ws}}=0.5$ and $U=8$.
First, we see that the wire spectral functions $A_{\text w}(\omega,k_x)$ in Figs.~\ref{fig:qmc_U8}(a) and (b)
are very similar despite the significant differences in the substrate spectral functions
$A_{\text s}(\omega,k_x)$ shown in Figs.~\ref{fig:qmc_U8}(c) and (d).
This confirms that the three-leg NLM can provide a good approximation of the wire properties
in the full 3D wire-substrate model.
The wire spectral functions closely resemble those of 1D Mott insulators~\cite{ben07,aich04,jec08b,racz15}.
A gap is clearly visible in Figs.~\ref{fig:qmc_U8}(a) and (b) and its size
agrees with the DMRG results of Fig.~\ref{fig:Ec_vs_U} within the numerical accuracy. The substrate band gap is also clearly seen in the substrate 
spectral functions $A_{\text s}(\omega,k_x)$ in Figs.~\ref{fig:qmc_U8}(c) and (d).
The gap in the wire spectral function is smaller than the effective substrate gap
$\Delta_{\text{s}}(N_{\text{leg}}=3) \approx 10$ of the three-leg NLM but quite close
to the true band gap $\Delta_{\text{s}} \approx 2$  of the 3D substrate.
Finally, Figs.~\ref{fig:qmc_U8}(a) and (c) reveal that
the spectral weight for the lowest single-particle excitations (i.e., for small $\vert \omega \vert$) 
of the three-leg NLM is concentrated exclusively in the wire. This confirms that these excitations are localized in the wire
in this model for $U < U_{{c}}$, as suggested by the spin and charge densities of the single-particle
excitations in Fig.~\ref{fig:densitiesHF}.

\begin{figure}
\includegraphics[width=0.5\textwidth]{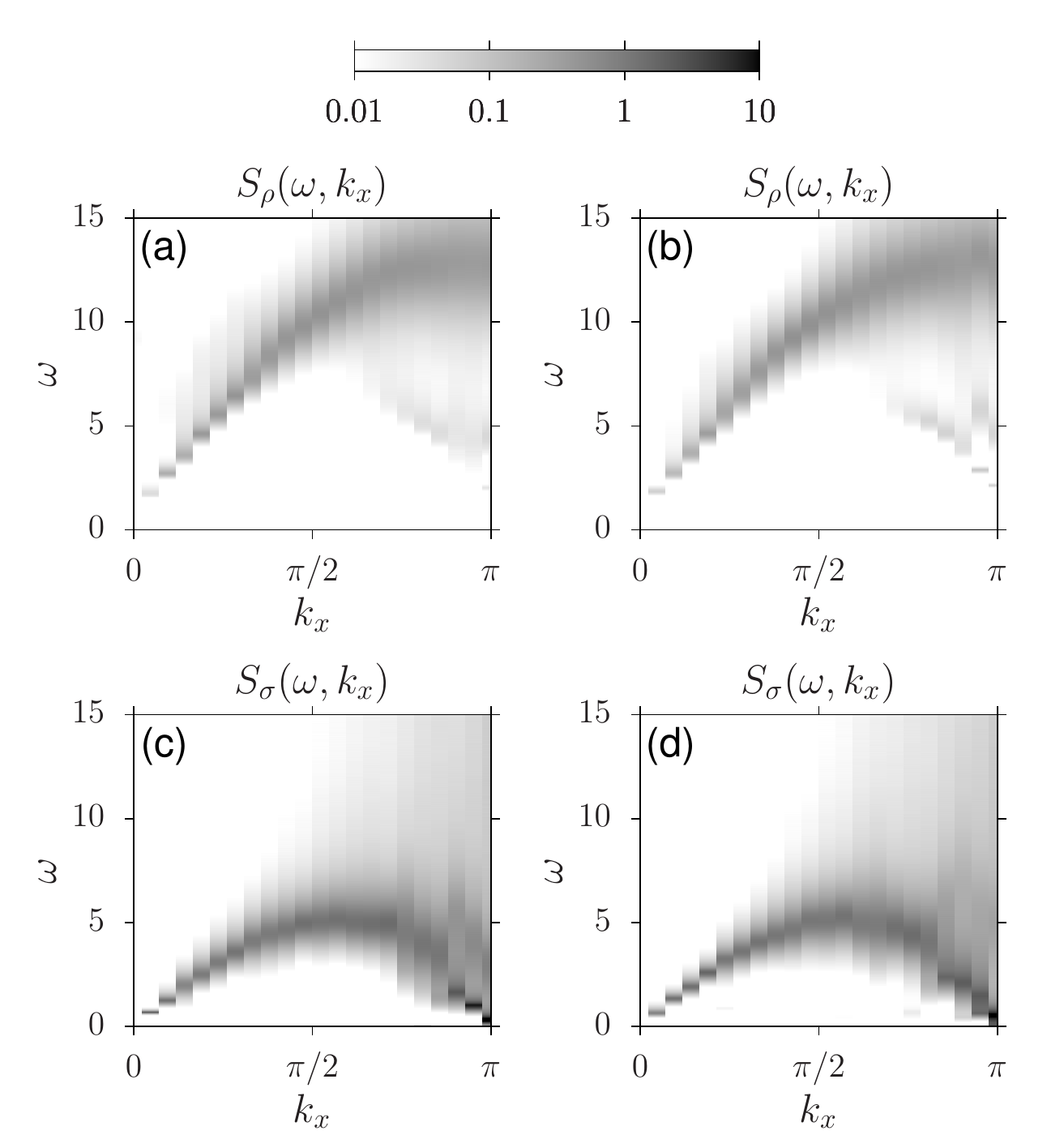}
\caption{\label{fig:qmc_U8b}
CT-INT results for the dynamic charge structure factor $S_{\rho}(\omega,k_x)$
[(a),(b)] and the dynamic spin structure factor $S_{\sigma}(\omega,k_x)$ [(c),(d)]
on the wire for the same parameters as in Fig.~\ref{fig:qmc_U8}.
Panels (a) and (c) show results for the three-leg NLM, panels (b) and (d)
for the 3D wire-substrate model.}
\end{figure}

Figure~\ref{fig:qmc_U8b} shows the charge and spin structure factors
of the wire for the same parameters as in Fig.~\ref{fig:qmc_U8}.
These spectra are very similar for the three-leg NLM and the 3D wire-substrate model, which again supports
the validity of the NLM approximation for the 1D physics occurring in the 3D wire-substrate model.
The low-energy features seen in $S_{\rho}(\omega,k_x)$ and $S_{\sigma}(\omega,k_x)$
resemble the ones found in 1D Mott insulators with gapless spin excitations~\cite{ben07,noce16b,racz15,pere12}.
The slope of the main feature in the spin structure factor for $k_x \rightarrow 0$  agrees
with our DMRG results for the velocity of spin excitations.

The spectral properties obtained with the CT-INT method are rather similar for 
other values of $U$. However, for weaker interactions $U$, we find 
that the spectral weight of the lowest  single-particle excitations  is concentrated mostly
on the wire not only for the NLM but also for the 3D wire-substrate model. This indicates
that the localization of the low-energy single-particle excitations on the 1D wire subsystem
is not an artifact of the NLM but a feature of the 3D wire-substrate model
in this parameter regime.

\begin{figure}
\includegraphics[width=0.5\textwidth]{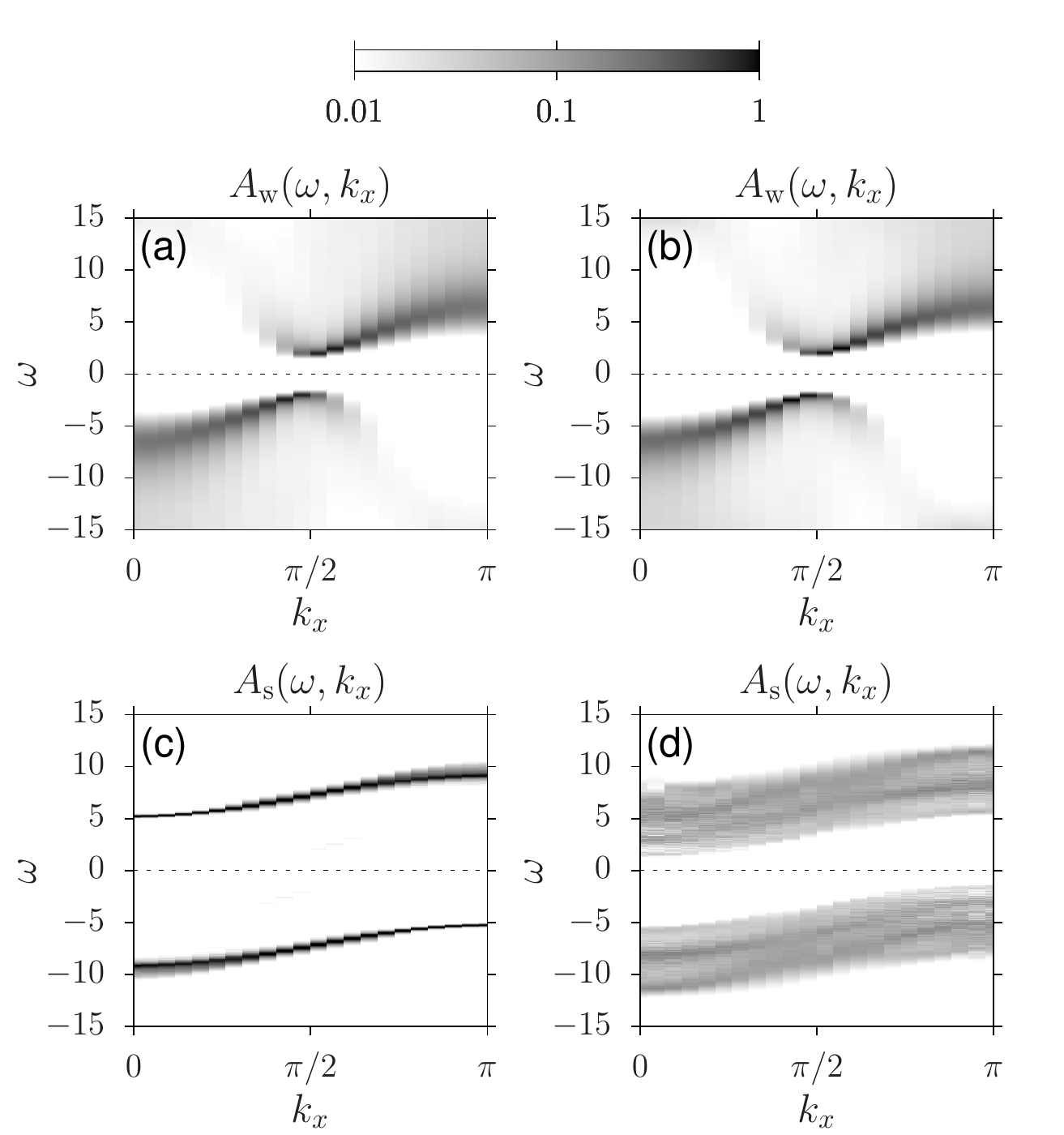}
\caption{\label{fig:qmc_U12}
CT-INT results for the spectral functions $A_{\text{w}}(\omega,k_x)$ [(a),(b)] and 
$A_{\text{s}}(\omega,k_x)$ [(c),(d)] for $U=12$, $t_{\text{ws}}=0.5$, $\beta=10$, and $L_x=42$.
The chemical potential was set to $\mu=0$, corresponding to
half-filling. Panels (a) and (c) show results for the three-leg NLM, panels
(b) and (d) for the 3D wire-substrate model ($L_y=42$,
$L_z=10$). 
}
\end{figure}

\begin{figure}
\includegraphics[width=0.5\textwidth]{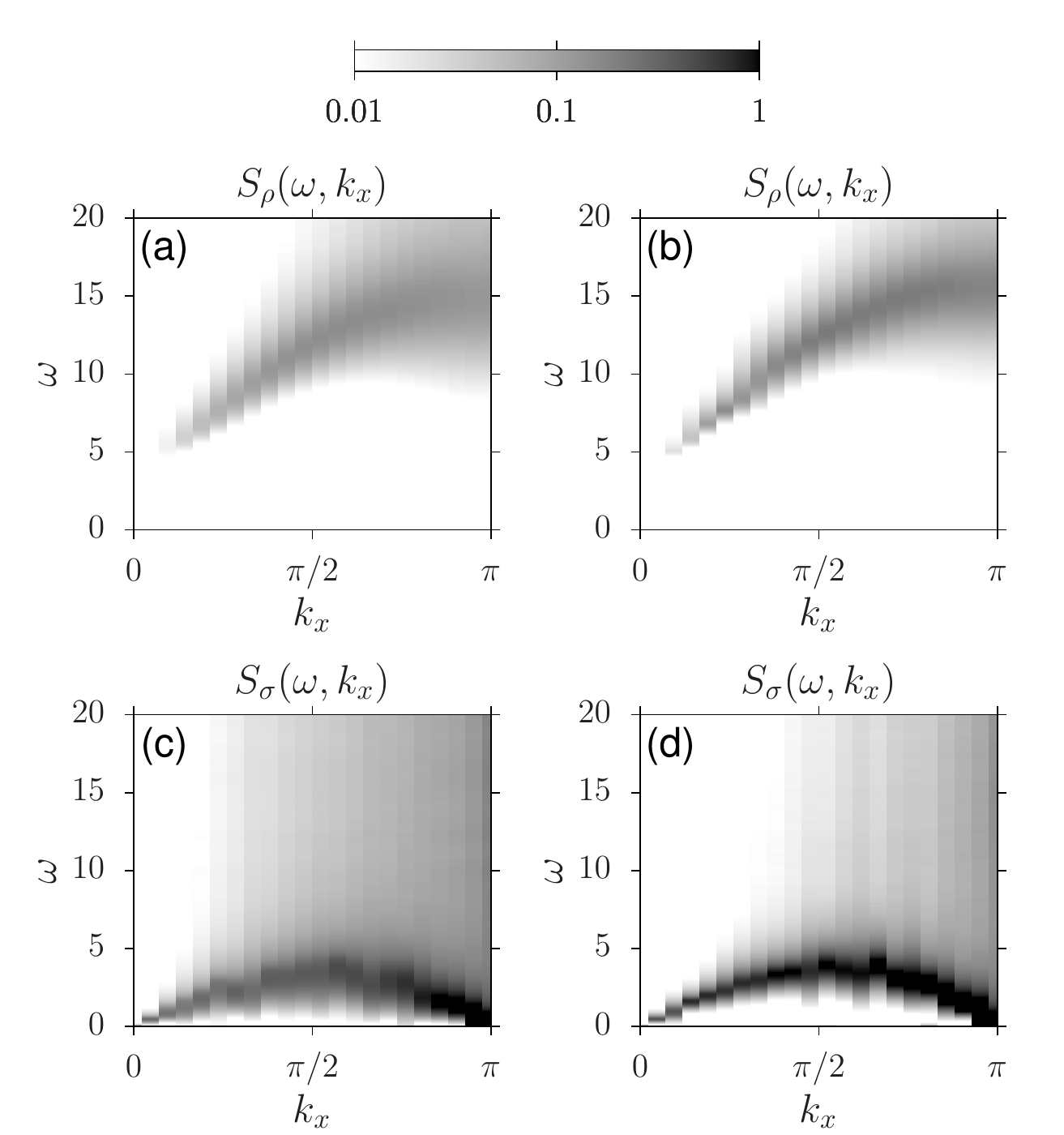}
\caption{\label{fig:qmc_U12b}
CT-INT results for the dynamic charge structure factor $S_{\rho}(\omega,k_x)$
[(a),(b)] and the dynamic spin structure factor $S_{\sigma}(\omega,k_x)$ [(c),(d)]
on the wire for the same parameters as in Fig.~\ref{fig:qmc_U12}.
Panels (a) and (c) show results for the three-leg NLM, panels (b) and (d) 
for the 3D wire-substrate model. }
\end{figure}

Figures~\ref{fig:qmc_U12} and~\ref{fig:qmc_U12b} show the spectral functions
and dynamic structure factors for $U=12$ 
(above the estimated critical coupling for the charge gap saturation 
$U_{c}\approx 9$ of the full 3D wire-substrate model at $t_{\text{ws}}=0.5$,
see below). We again see that the spectral properties of the wire are very similar
in the three-leg NLM and the 3D wire-substrate model.
In contrast to $U=8$,
Figs.~\ref{fig:qmc_U12}(b) and (d) reveal that
the gap in the single-particle spectral function for the wire ($\Delta \omega \approx  4$) is
comparable to the Mott gap of the 1D Hubbard chain and thus significantly
larger than the substrate band gap of the 3D wire-substrate model ($\Delta_\text{s} \approx 2$).
Thus low-energy single-particle excitations now involve the valence and conduction bands
and are delocalized in the substrate.   
This agrees qualitatively with our DMRG results for the three-leg NLM
above its critical coupling $U_{c}\approx 20$, i.e., the charge gap saturation
in Fig.~\ref{fig:Ec_vs_U} as well as
the spin and charge densities of single-particle excitations in
Fig.~\ref{fig:densitiesHF}.

Figure~\ref{fig:qmc_U12b}(b) shows that the dynamic charge structure factor of the wire
has no spectral weight at energies between the substrate band gap at $\omega \approx 2$ 
and the single-particle gap of the wire at $\omega \approx 4$.
This suggests that the system still has charge excitations localized on the wire 
but only at high energy, i.e., above the Mott gap. 
Finally, the spin structure factor of the 3D wire-substrate model in Fig.~\ref{fig:qmc_U12b}(d)
confirms the existence of gapless spin excitations localized on the wire
even though the lowest single-particle excitations seem to be in the band-insulating substrate. 
This again agrees qualitatively with the DMRG results for the three-leg NLM above its critical 
coupling $U_{c}\approx 20$, in particular the vanishing of the spin gap illustrated
in Fig.~\ref{fig:Es_vs_L} and the density distribution for spin excitations shown in Fig.~\ref{fig:densitiesHF}.

We conclude that the three-leg NLM describes a quasi-1D Mott insulator
with gapless spin excitations for weak Hubbard interaction $U<U_{{c}}$,
at least for $t_{\text{ws}}\lesssim 4$
and $U\gtrsim 4$.
Increasing $t_{\text{ws}}$ reduces the charge gap, and thus the effective repulsion between charges,
but also reduces the spin velocity and thus the effective spin exchange coupling
compared to an isolated Hubbard chain. 
Thus, we cannot explain the properties of this phase
by a 1D model with an effective onsite Hubbard interaction only
because, in the latter, the effective exchange coupling decreases as the charge
gap increases~\cite{essler05}.

For strong Hubbard interactions $U > U_{c}$ we find a transition to a band insulator.
The lowest charge and single-particle excitations are then transitions between 
the valence and conduction bands of the NLM representations, as shown
by their energy in Fig.~\ref{fig:Ec_vs_U} and their density variations in Fig.~\ref{fig:densitiesHF}.
In addition, the spins of the electrons localized on the wire leg represent magnetic impurities
embedded in the band insulator. They form an effective Heisenberg chain with the gapless excitations seen in Fig.~\ref{fig:Es_vs_L}. 
Unfortunately, the spectral properties in this regime are not accessible by
the CT-INT method because of the large expansion orders for $U> U_{{c}} \approx 20$.

The mechanism of the transition can be understood for weak hybridization $t_{\text{ws}}$
starting from the noninteracting limit discussed in \cite{paper1}.
The effective electron-electron interaction in the wire opens a Mott gap in the middle of the wire band
as seen in Figs.~\ref{fig:qmc_U8}(a) and (b).
This gap grows with increasing $U$ until it reaches  
the effective gap between the substrate bands.
For even larger $U$, the band gap is smaller than the Mott gap and the nature of the elementary excitations changes
from holons and spinons in a quasi-1D Mott insulator to electrons and holes in a band insulator.
Indeed, for weak hybridization $t_{\text{ws}}$, the effective Mott gap is almost equal to the gap of the 1D Hubbard chain
and the transition occurs exactly when this gap equals the effective band gap of the noninteracting NLM,
$\Delta_{\text{s}}(N_{\text{leg}}=3)\approx 10$ or $\Delta_{\text{s}}(N_{\text{leg}}=7)\approx 5.5$, 
see Fig.~\ref{fig:Ec_vs_U}.
It is remarkable that this scenario remains qualitatively unchanged up to at least $t_{\text{ws}}=4$.

As discussed above, the effective substrate gap $\Delta_{\text{s}}(N_{\text{leg}})$ is
considerably reduced upon increasing $N_\text{leg}$ until it reaches the value of the true substrate band gap.
Accordingly, $U_{{c}}$ decreases for higher numbers of legs, as
seen in Fig.~\ref{fig:Ec_vs_U}.
Although we cannot simulate large enough correlated NLMs to observe the convergence of $U_{{c}}$ 
with $N_{\text{leg}}$, we expect that it remains finite in the full 3D wire-substrate system
with a finite band gap.
Using the criterion discussed above for weak hybridization $t_{\text{ws}}$
(i.e., the 1D Hubbard gap equals the substrate band gap $\Delta_{\text{s}}=2$), we can estimate from Fig.~\ref{fig:Ec_vs_U} that 
$U_{{c}} \approx 9$ in the full 3D wire-substrate system for small $t_{\text{ws}}$
and that $U_{{c}}$ becomes larger for stronger hybridization.
This interpretation agrees perfectly with the spectral properties 
computed with the CT-INT method for the 3D wire-substrate model. 
In particular, the single-particle spectral functions in Figs.~\ref{fig:qmc_U12}(b) and (d)
demonstrate that this system is a band insulator for $U=12$ while the corresponding structure factors
in Figs.~\ref{fig:qmc_U12b}(b) and (d) confirm that a 1D subsystem with high-energy charge excitations
($\Delta \omega \agt 4$)
but gapless spin excitations (like a spin-$1/2$ Heisenberg chain) is embedded in that band insulator.
Therefore, we think that the transition from the quasi-1D Mott insulator to a band insulator is not 
an artifact of the three-leg NLM but a feature of the correlated 3D
wire-substrate model that is qualitatively reproduced by the approximate NLM.

It is known~\cite{essler05} that the half-filled 1D Hubbard model undergoes a phase transition
from a metallic Fermi gas at $U=0$ to a Mott insulator for $U>0$. Although we
cannot distinguish these phases numerically for very weak  $U$, we expect that a similar transition occurs in the NLM and hence in the 3D wire-substrate model.
However, this should be confirmed by methods that are better suited for the weakly interacting regime, such as 
field-theoretical approaches~\cite{sol79,giamarchi07,Schoenhammer,Gogolin,Tsvelik} for the three-leg NLM.

\subsection{Metallic wire}

\begin{figure}
\includegraphics[width=0.4\textwidth]{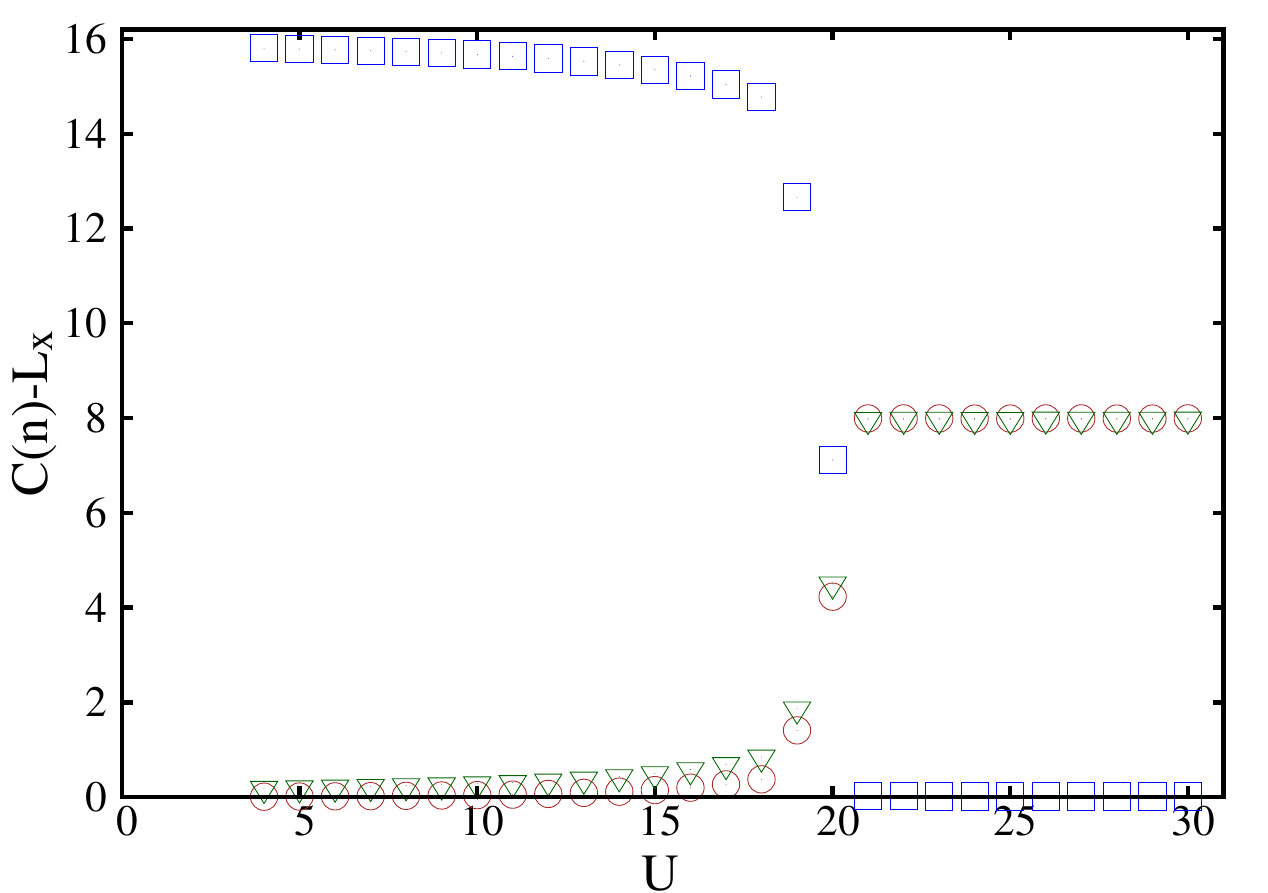}
\caption{\label{figchrgspindens} (Color online)
DMRG results for the difference between the total charge away from half-filling $C(n)$ and at half-filling $C(n)=L_x$ 
in the wire leg (squares) as well as in the first (circles) and second (triangles) substrate legs
as a function of the interaction $U$. The data are for a three-leg
NLM with length $L_x=128$, $t_{\text{ws}}=0.5$ and a wire doping $y_{\text w}=12.5\%$.
}
\end{figure}

We now turn to the discussion of doped systems. For DMRG calculations we focused on the case
of $L_x/8$ added electrons, corresponding to a wire doping of $y_{\text w} = 12.5\%$.
Removing electrons gives similar results due to electron-hole symmetry.
Similarly, in the QMC simulations, the chemical potential was tuned
to obtain $y_{\text w} \approx 12.5\%$. Due to the
different size of the substrate band gap, different chemical potentials were
required for the three-leg NLM and the 3D wire-substrate model. The
corresponding values of $\mu$ are given in the figure captions.

For $t_{\text{ws}}=0$ (and $U < \Delta_\text{s}$), the wire corresponds to a doped Hubbard chain.
The ground state in this case is a 1D conductor~\cite{essler05} with the low-energy properties of a Luttinger liquid~\cite{giamarchi07}.
We expect these systems to be quasi-1D conductors also for $t_{\text{ws}}\neq 0$ and
to yield information that could be relevant for understanding the numerous metallic atomic wires
studied experimentally~\cite{sni10,blu11,blum12,naka12,park14,jong16,ohts15,yaji13,yaji16,Tegenkamp2005,Wanke2011}.

Figure~\ref{figchrgspindens} shows the variations of the charge distributions $C(n)$
relative to half-filling as a function of the
interaction $U$. For weak $U$, most of the added charges go on the wire leg ($y_{\text{eff}} \approx y_{\text w}$) 
while for strong $U$ they go on the substrate legs ($y_{\text{eff}} \ll y_{\text w}$). 
The crossover---which seems to be continuous but abrupt---occurs close to 
the critical $U_c\approx 20$ found at half-filling and the charge distribution
is consistent with the transition from a Mott to a band insulator observed at half-filling.
The added electrons occupy states corresponding to the lowest excited states, i.e., in the upper
Hubbard band localized on the wire for $U \alt U_c$ but in the conduction band localized on the substrate legs
for $U \agt U_c$.

\begin{figure}
\includegraphics[width=0.4\textwidth]{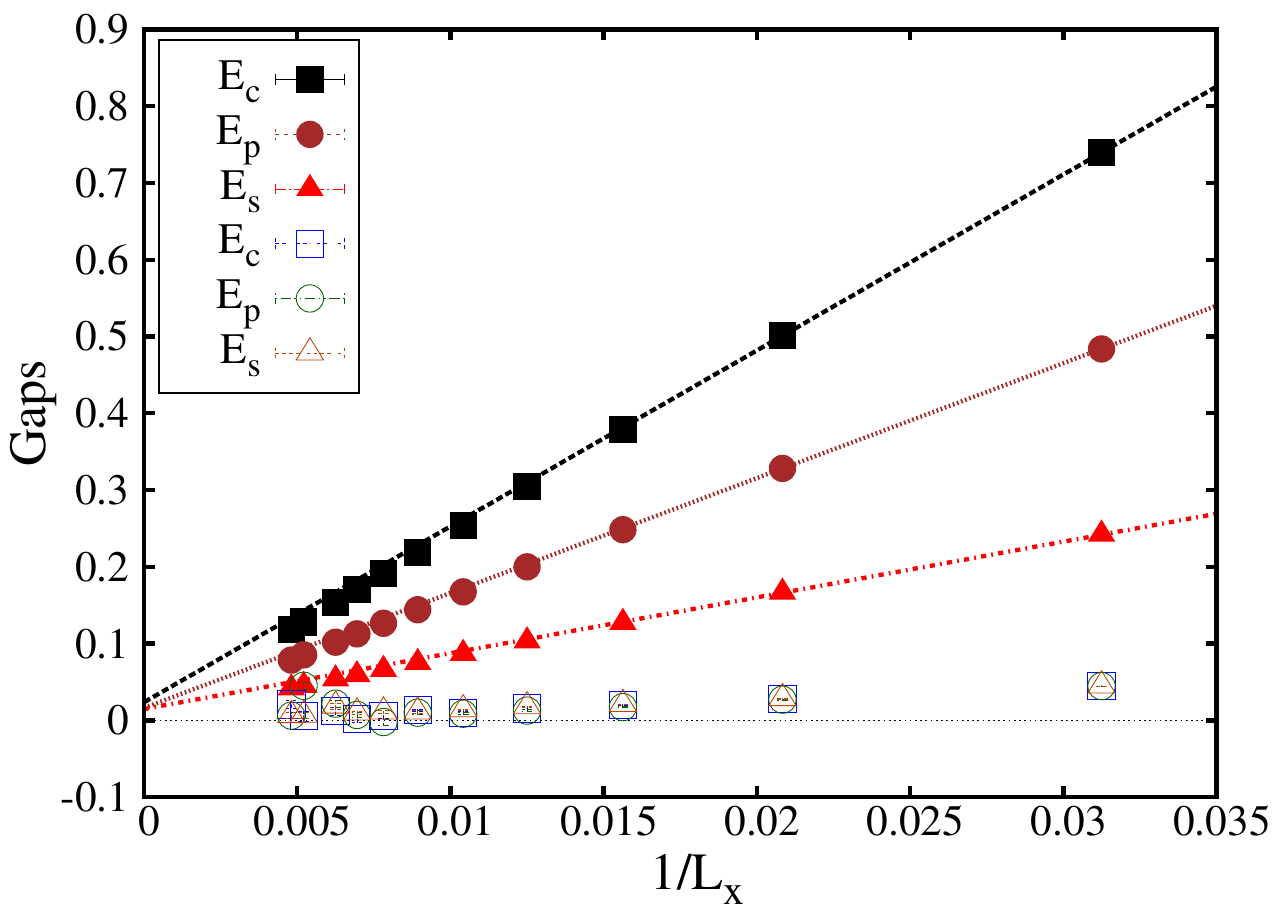}
\caption{\label{fig:chrgspingaps} (Color online)
DMRG results for the charge $(E_{\text{c}})$, spin $(E_{\text{s}})$, and single-particle $(E_{\text{p}})$ gaps of the three-leg NLM away 
from half-filling ($y_{\text{w}}=12.5\%$). Here, $t_{\text{ws}}=0.5$ and
$U=16$ (filled symbols) and $U=24$ (open symbols), respectively.
Slanting lines correspond to linear fits.
}
\end{figure}

Away from half-filling, the charge, spin and single-particle gaps vanish in the thermodynamic limit
for any $U \geq 0$. However, the doped Mott and band insulating phases exhibit significantly different finite-size
effects, as illustrated in Fig.~\ref{fig:chrgspingaps} for $U=16$ and $U=24$.
For $U \alt U_c$, the finite-size gaps vanish linearly with $1/L_x$. The 
velocities defined by Eq.~(\ref{eq:velocities}) (i.e., the fitted slopes in Fig.~\ref{fig:chrgspingaps}) 
are larger for charge excitations ($v_{\text{c}}$) than for spin excitations 
($v_{\text{s}}$) and about the average of $v_{\text{c}}$ and $v_{\text{s}}$
for single-particle excitations ($v_{\text{p}}$).
For $U \agt U_c$, the charge, spin and single-particle gaps are equal (within the DMRG errors)
and much smaller than for weak interactions. The relative DMRG errors for these gaps are too large
to accurately determine their scaling with $1/L_x$.

\begin{figure}
\includegraphics[width=0.4\textwidth]{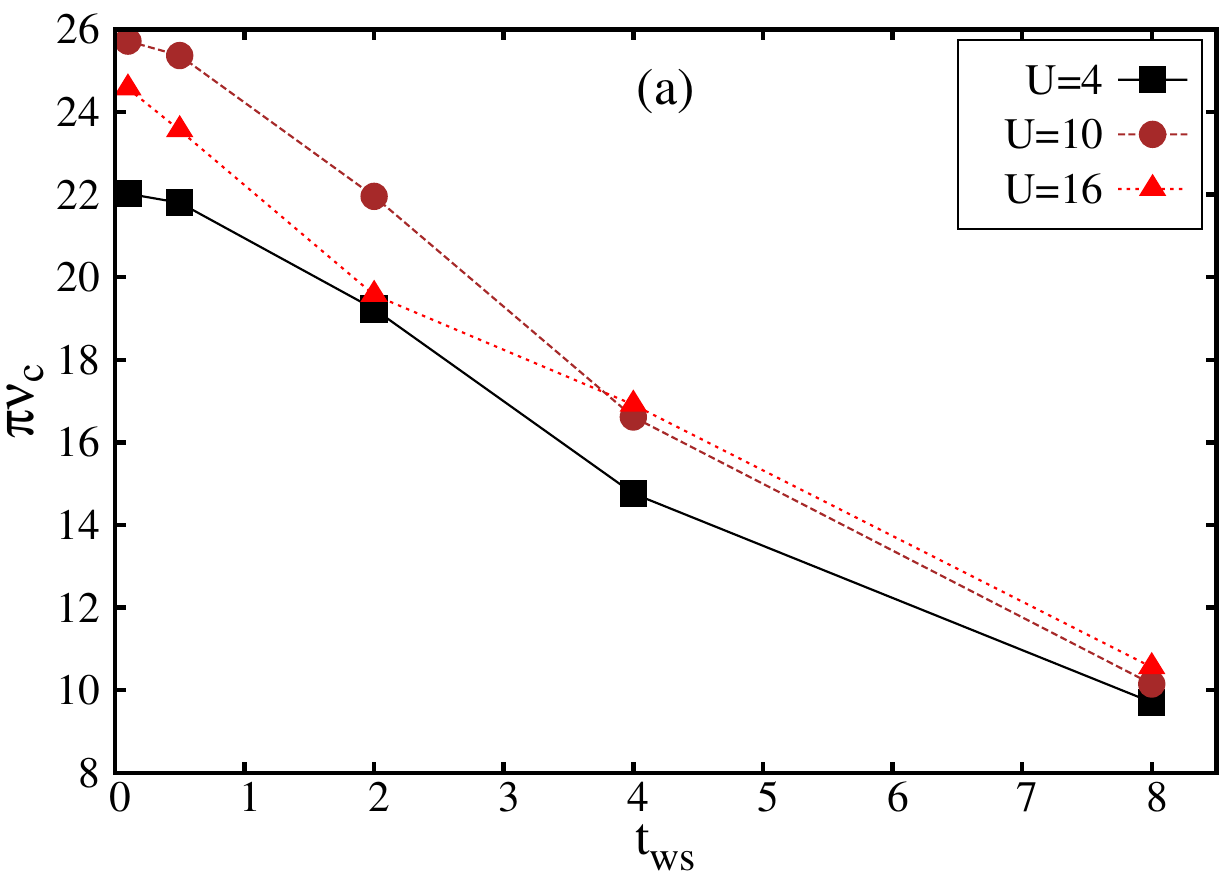}
\includegraphics[width=0.4\textwidth]{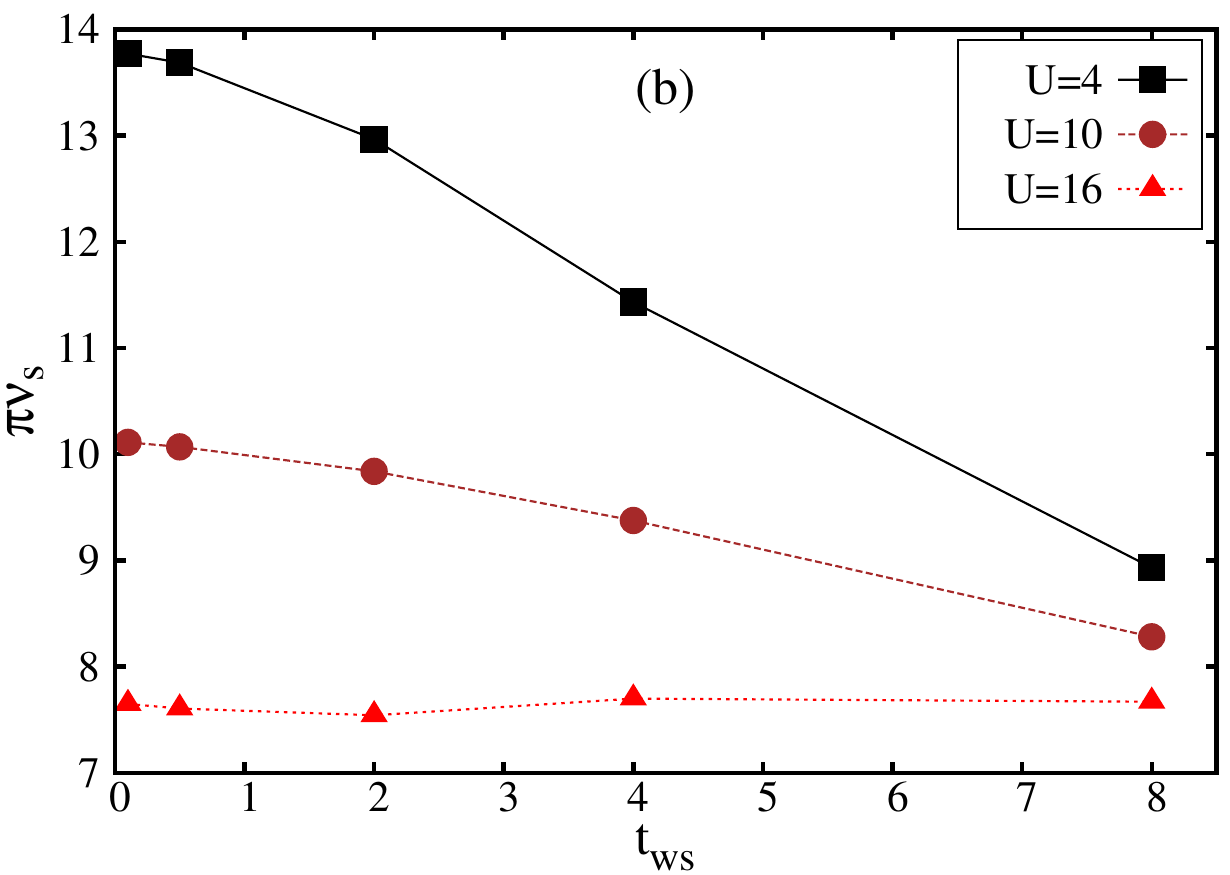}
\caption{\label{fig:velocities} (Color online)
DMRG results for (a) charge and (b) spin velocities of the doped three-leg NLM ($y_{\text w}=12.5\%$)
as a function of the wire-substrate hybridization strength $t_{\text{ws}}$ for  three values of
the Hubbard interaction $U$. The results were obtained from the
finite-size scaling of the corresponding excitations gaps [cf. Eq.~(\ref{eq:velocities})].
}
\end{figure}

We systematically investigated the velocities of elementary excitations in the weak-coupling phase ($U \alt U_c$) of the three-leg NLM. As
expected, the results approach those for the 1D Hubbard model when $t_{\text{ws}}$ becomes very
small. Although fitting the finite-size DMRG gaps introduces uncertainties,
we can recognize two trends in Fig.~\ref{fig:velocities}. First, the velocities decrease with increasing
wire-substrate hybridization $t_{\text{ws}}$. Second, as observed in
the 1D Hubbard model, spin velocities are significantly
reduced upon increasing $U$ whereas charge velocities are only weakly
affected. Similar to half-filling, we verified that these velocities do not differ
significantly in wider NLMs with up to $N_{\text{leg}}=7$.
The different charge and spin velocities are a signature 
of dynamic spin-charge separation typical of the Luttinger liquid state
obtained by doping a 1D Mott insulator. However, for $t_{\text{ws}} > 0.5$,
the dependence of the velocities on $U$ is different from a
doped 1D Hubbard model and thus cannot be captured by an effective onsite interaction only.

\begin{figure}
\includegraphics[width=0.4\textwidth]{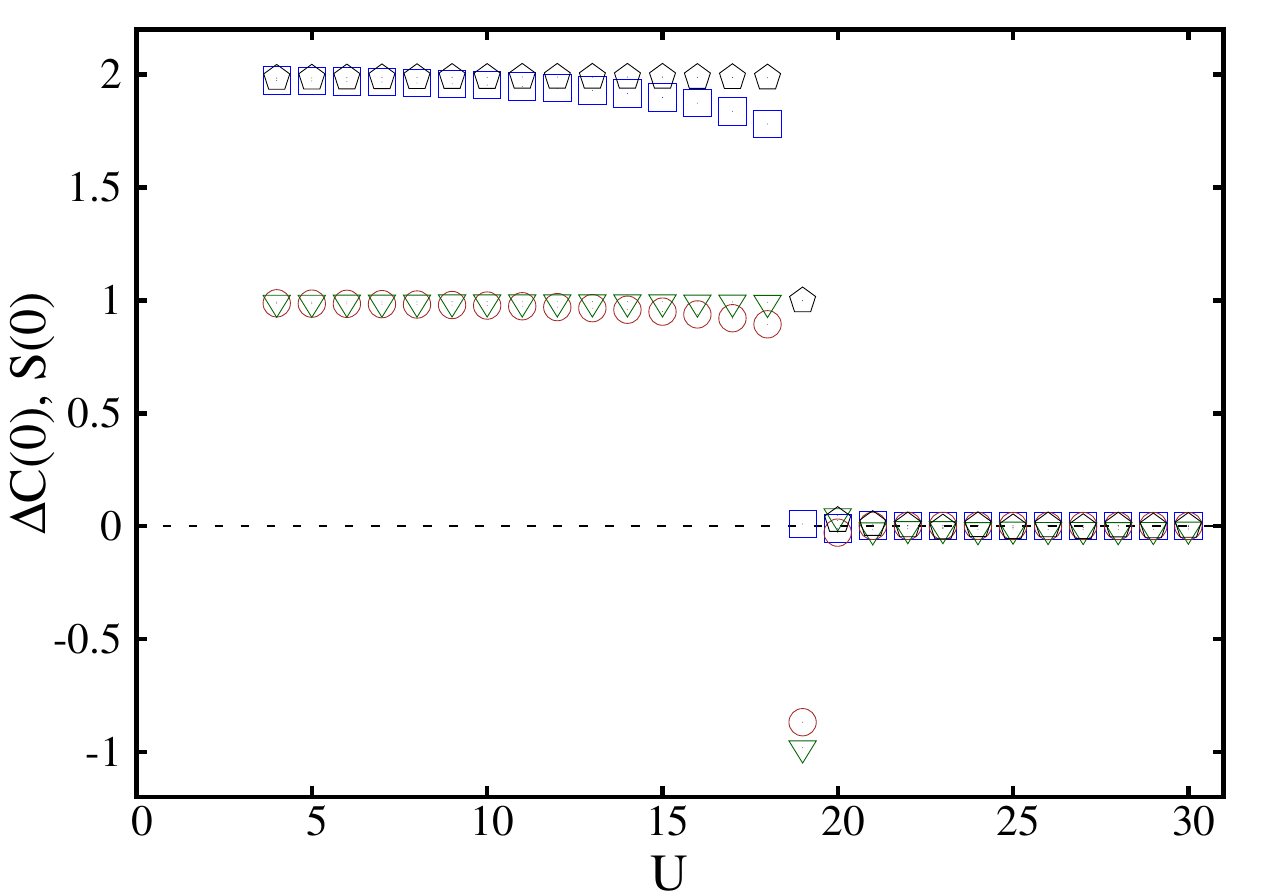}
\caption{\label{fig:densitiesAFHF} (Color online)
DMRG results for the variations of charge [$\Delta C(0)$] and spin [$\Delta S(0)=S(0)$]
on the wire leg as a function of $U$ for the doped
three-leg NLM ($y_{\text{w}}=12.5\%$). The different
symbols correspond to the lowest charge (squares), spin (pentagons), and
single-particle (circles and triangles, respectively) excitations.
}
\end{figure}

Additionally, we investigated the location of the lowest charge and spin
excitations in the doped three-leg NLM.
Figure~\ref{fig:densitiesAFHF} shows the variations of charge and spin 
distributions between excited states and the ground state, similar to Fig.~\ref{fig:densitiesHF} for half-filling. 
In the weak-coupling regime $U \alt U_c$, charge, spin and single-particle
excitations are almost entirely localized on the wire leg
for moderate wire-substrate hybridization strengths (e.g.,
$t_{\text{ws}}=0.5$), similar to the half-filled case.
(For larger $t_{\text{ws}}$, charge and spin excitations can be partially localized
on both wire and substrate legs and the dependence on $U \alt U_c$ is more
complex.) In contrast, in the strong-coupling regime $U \agt U_c$, 
low-energy excitations are predominantly localized on the substrate
legs. Again, we have checked that these uneven distributions persist in wider NLMs with up to $N_{\text{leg}}=7$ legs. 

\begin{figure}
\includegraphics[width=0.4\textwidth]{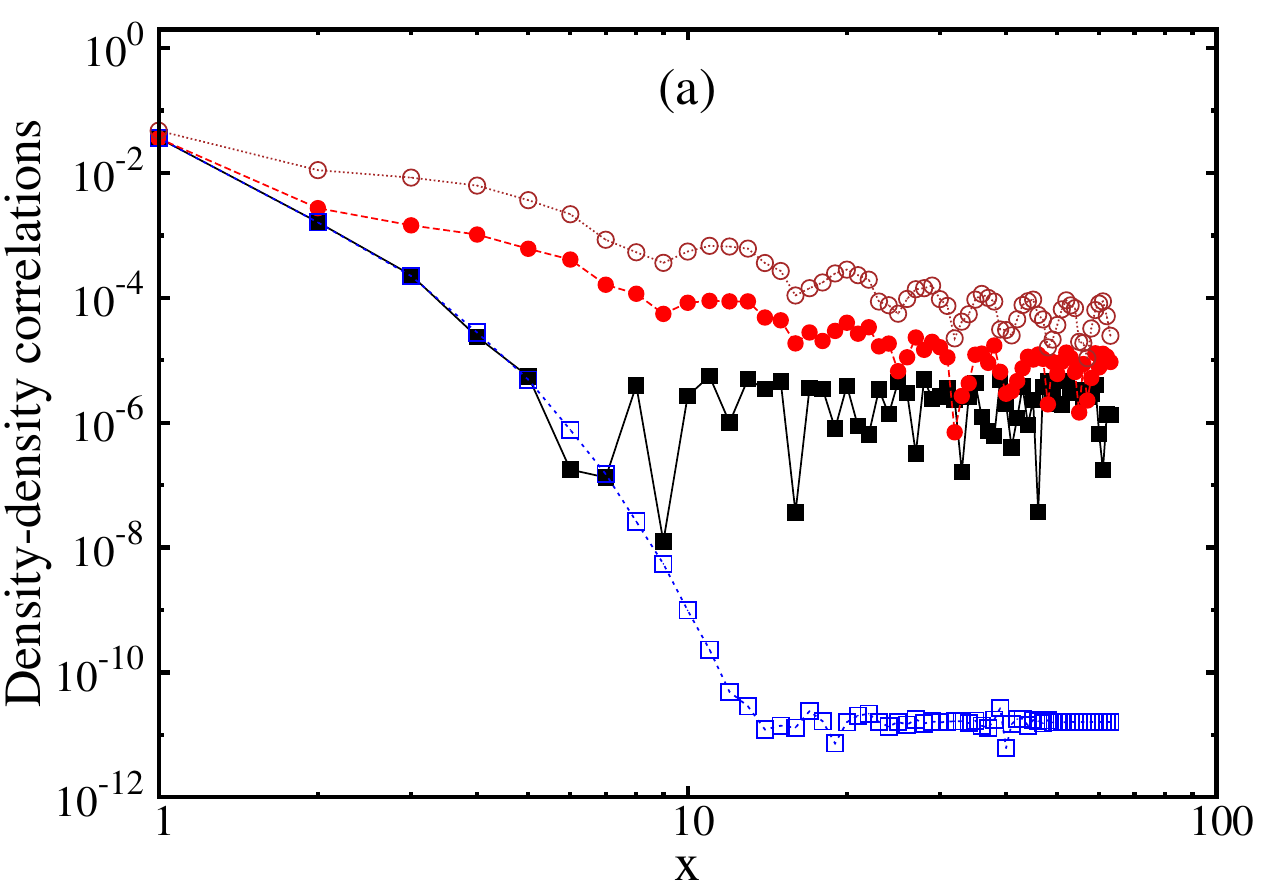}
\includegraphics[width=0.4\textwidth]{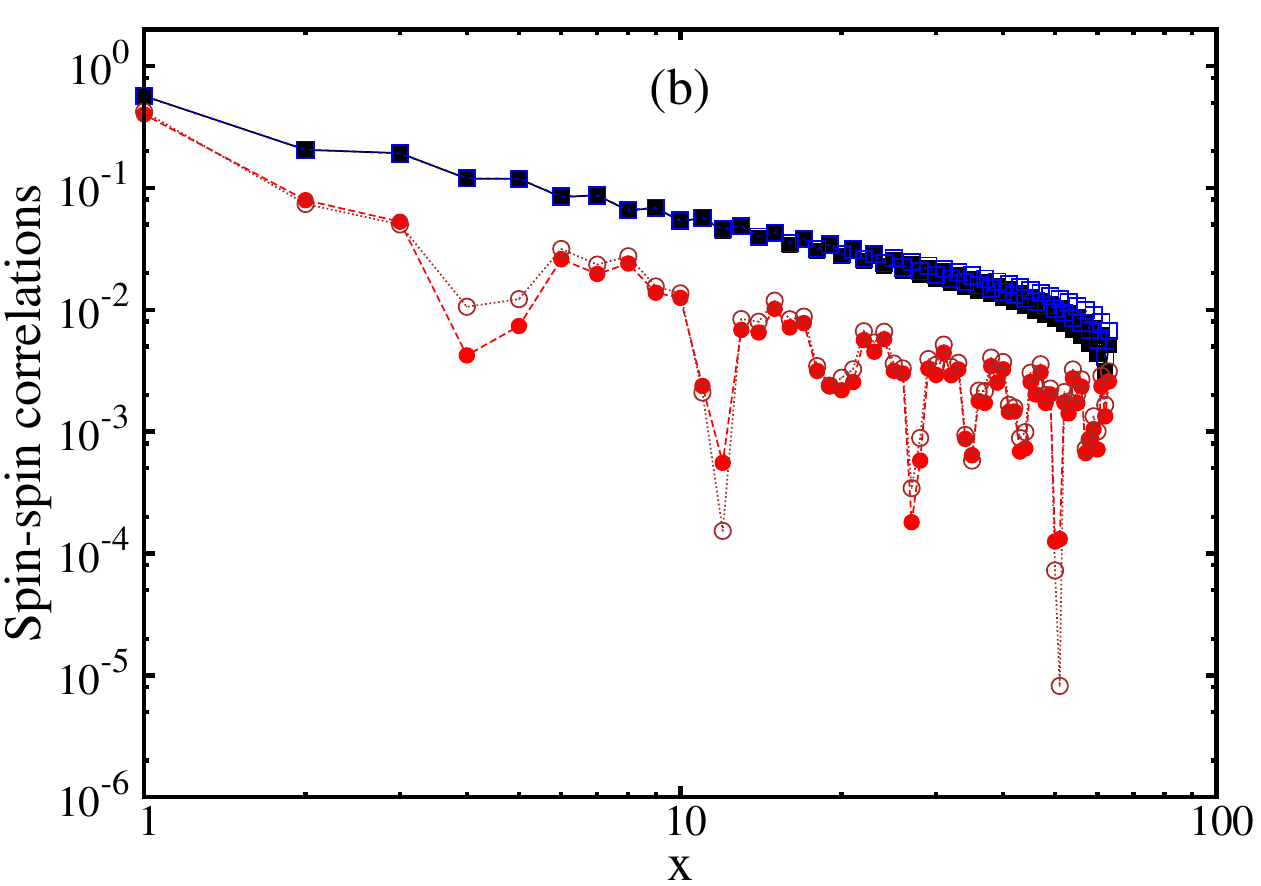}
\caption{\label{fig:correl} (Color online)
DMRG results for the absolute values of the (a) charge and (b) spin
correlation functions on the wire leg for the doped three-leg NLM 
($y_{\text{w}}=12.5\%$). Here, $t_{\text{ws}}=0.5$ (filled squares) and
$t_{\text{ws}}=2$ (filled circles), respectively. Also shown are results for 
the 1D Hubbard model with $t_{\text{w}}=3$ at half-filling (open squares) and
at $12.5\%$ doping (open circles). All results are for $U=24$. 
}
\end{figure}

Finally, it should be noted that in the strong-coupling phase the wire is still present as a quasi-1D correlated impurity 
embedded in the substrate, similar to half-filling. This is clearly visible from the behavior of charge and 
spin density correlations along the wire. The charge correlation function for the wire is defined by
\begin{eqnarray}
F_{\text{c}}(x-x') & = & \left \langle \sum_{\sigma}n_{wx\sigma} \sum_{\sigma'}n_{wx'\sigma'}
 \right \rangle \\
&&
- \left \langle \sum_{\sigma} n_{wx\sigma} \right \rangle 
\left \langle \sum_{\sigma'} n_{wx'\sigma'}  \right \rangle 
\nonumber
\end{eqnarray}
while the spin correlation function is 
\begin{equation}
F_{\text{s}}(x-x')= \left \langle \sum_{\sigma} \sigma \, n_{wx\sigma} 
\sum_{\sigma'} \sigma' \, n_{wx'\sigma'} \right \rangle 
\end{equation}
with $n_{wx\sigma}  = c^{\dag}_{{\text w}x\sigma}c^{\phantom{\dag}}_{\text{w}x\sigma}=g^{\dag}_{x0\sigma}g^{\phantom{\dag}}_{x0\sigma}$.
Here, expectation values are with respect to the ground state $\vert \psi_{\text{GS}} \rangle$.
These correlation functions are shown in Fig.~\ref{fig:correl} for $U=24$ in the strong-coupling phase 
($t_{\text{ws}}=0.5$, $U_c \approx 20$) and in the weak-coupling phase ($t_{\text{ws}}=2.0$, $U_c \approx 32$).
In the former case, doped particles populate the substrate, as discussed above,  
whereas the wire sites are still occupied by one electron on average---as in
the half-filled Hubbard model---despite the doping of the three-leg NLM. 
Accordingly, Fig.~\ref{fig:correl}(a) shows that charge density correlations in the wire
decay exponentially for short distances $x$
in quantitative agreement with the half-filled Hubbard model with the same $U$. 
[The saturation of $F_{\text{c}}(x)$ at long distances is due to DMRG errors and
additional interference from the power-law correlations in the substrate legs.]
Similarly, Fig.~\ref{fig:correl}(b) shows that spin correlations in the wire decay
with a power-law with an exponent close to $-1$, in quantitative agreement with
the half-filled Hubbard model. 
In contrast, in the weak-coupling phase, doped particles populate the wire, resulting
in an average density different from one electron per wire site.
Correspondingly, the NLM exhibits a power-law decay of charge and spin density correlations,
in qualitative agreement with the behavior of these correlation functions in a Hubbard chain with 
a similar doping $\approx 12.5 \%$ (also shown in Fig.~\ref{fig:correl}).

\begin{figure}
\includegraphics[width=0.5\textwidth]{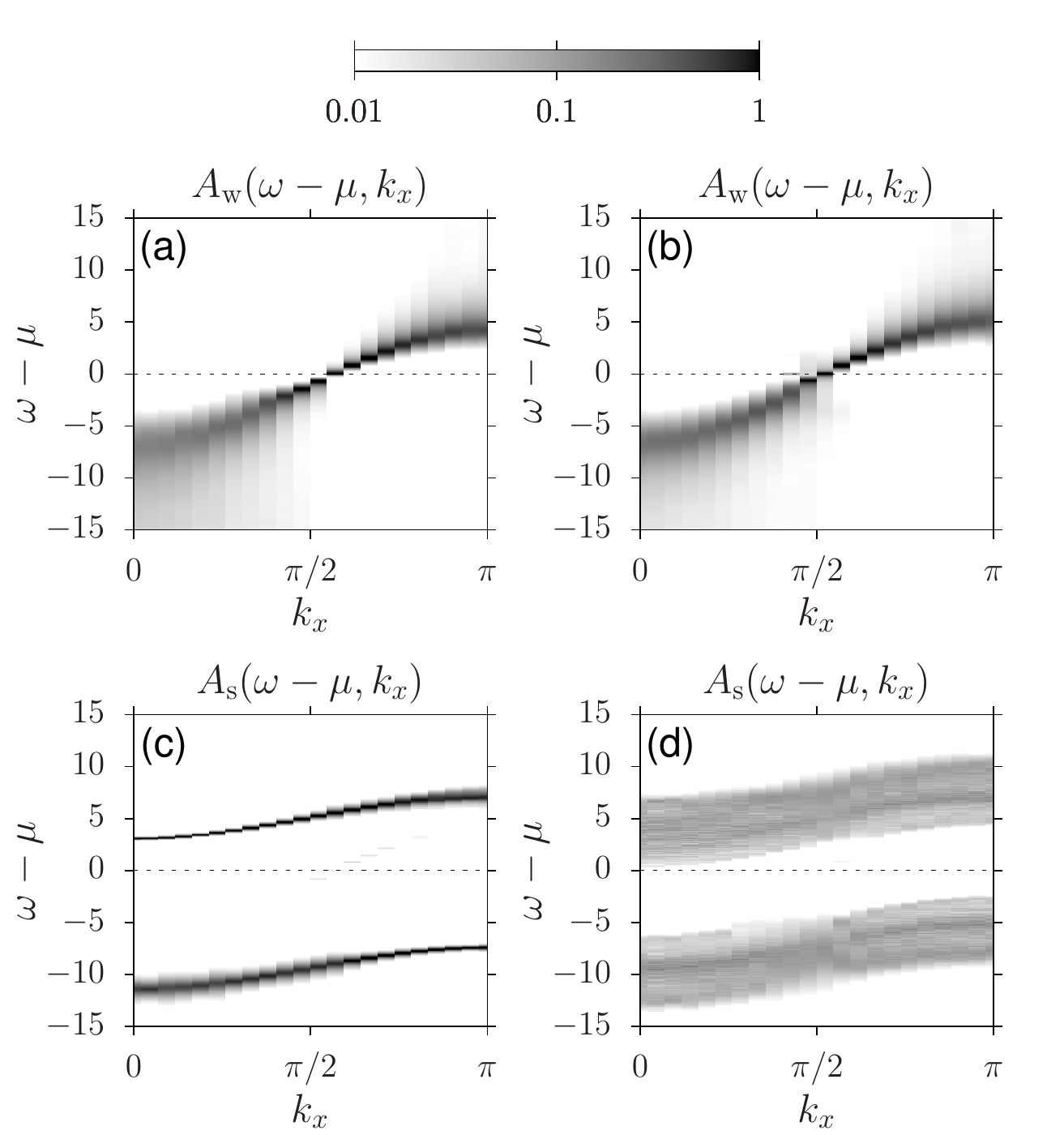}
\caption{\label{fig:qmc_U8_doped}
CT-INT results for the spectral functions $A_{\text{w}}(\omega,k_x)$ [(a),(b)] and 
$A_{\text{s}}(\omega,k_x)$ [(c),(d)] for $U=8$, $t_{\text{ws}}=0.5$, $\beta=15$, and $L_x=42$.
Panels (a) and (c) show results for the three-leg NLM, panels (b) and (d) for the 3D wire-substrate model ($L_y=42$, $L_z=10$).
The chemical potential was $\mu=2.1375$ for the NLM and  $\mu=0.99$ for the 3D
model, corresponding to a doping of $y_\text{w}\approx 12.5\%$.
}
\end{figure}

\begin{figure}
\includegraphics[width=0.5\textwidth]{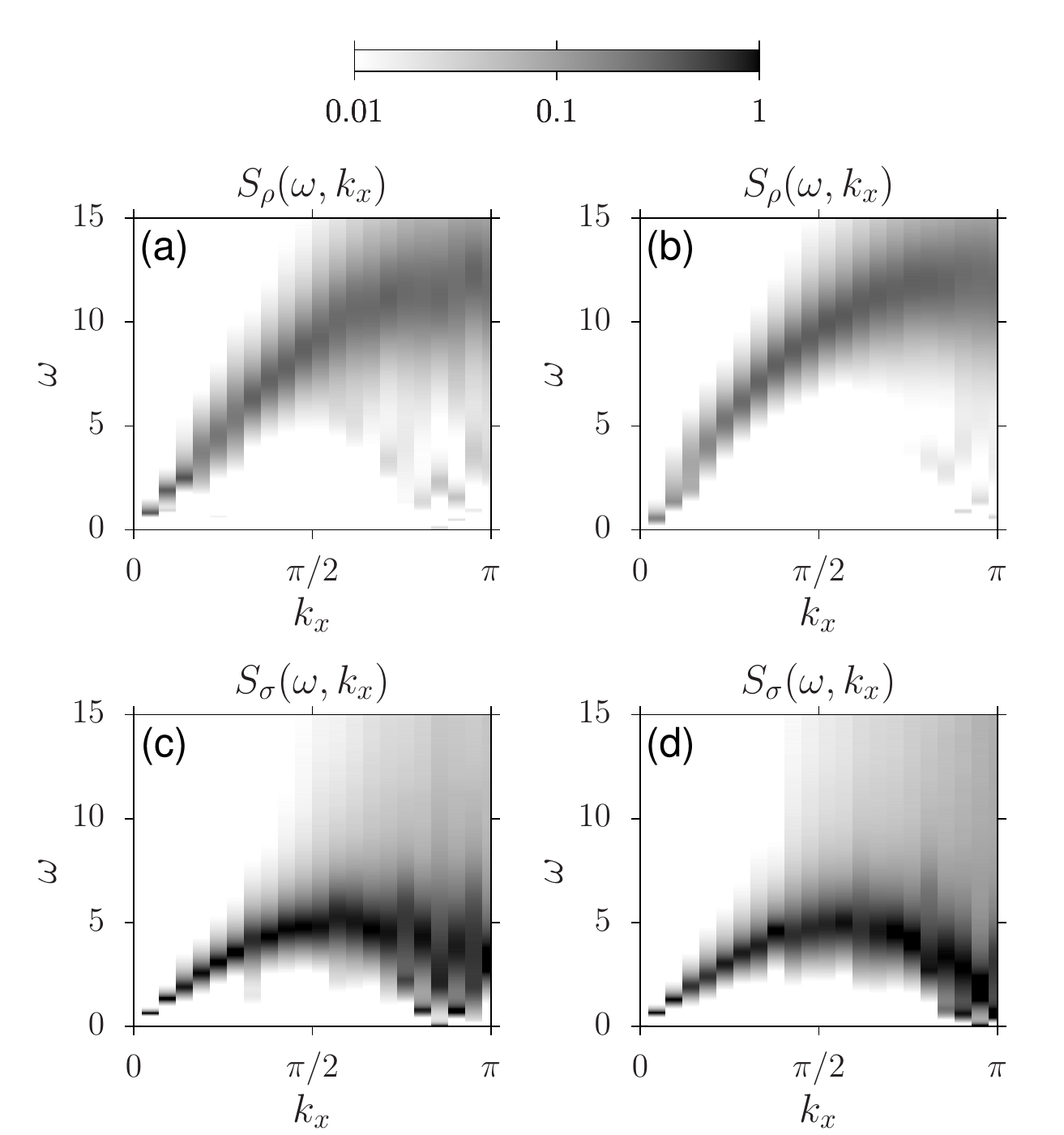}
\caption{\label{fig:qmc_U8_doped2}
CT-INT results for the dynamic charge structure factor $S_{\rho}(\omega,k_x)$
[(a),(b)] and the dynamic spin structure factor $S_{\sigma}(\omega,k_x)$ [(c),(d)]
on the wire for the same parameters as in Fig.~\ref{fig:qmc_U8_doped}.
Panels (a) and (c) show results for the three-leg NLM, panels (b) and (d) for
the 3D wire-substrate model.}
\end{figure}

Additional evidence for two distinct phases comes from the spectral properties calculated with the CT-INT method.
Figure~\ref{fig:qmc_U8_doped} shows the single-particle spectral functions of the three-leg NLM and the 3D wire-substrate
model for a finite wire doping. The model parameters are the same as in Fig.~\ref{fig:qmc_U8} for half-filling.
A Hubbard parameter $U=8$ puts the system in the Luttinger liquid region
according to the DMRG results.
The wire spectral functions are almost identical for the
three-leg NLM [Fig.~\ref{fig:qmc_U8_doped}(a)] and the 3D wire-substrate
model [Fig.~\ref{fig:qmc_U8_doped}(b)]. 
They are qualitatively similar
to those of the doped 1D Hubbard model~\cite{ben04,jec08,aben06} and
compatible with the field-theoretical predictions for Luttinger liquids~\cite{med92,mark16}.
In particular, they clearly show the presence of gapless single-particle excitations.
In contrast, Fig.~\ref{fig:qmc_U8_doped}(c) does not reveal any low-energy excitations in the substrate spectral function
of the three-leg NLM. In Fig.~\ref{fig:qmc_U8_doped}(d) the Fermi energy (i.e, $\omega=\mu$) still lies in the substrate band gap
but very close to the bottom of the conduction band
and the little spectral weight at $\omega=\mu$ is due to the finite temperature $\beta^{-1}$
used in the QMC simulations.
Therefore, the CT-INT single-particle spectral functions corroborate the existence of gapless low-energy excitations
localized in the wire predicted by the DMRG results.
Moreover, they confirm that the three-leg NLM can describe such excitations as well as
the 3D wire-substrate model.

The corresponding dynamic charge and spin structure factors of the wire are shown in Fig.~\ref{fig:qmc_U8_doped2}.
Again we see that the spectra are similar for the three-leg NLM 
and the 3D wire-substrate model.
The structure factors resemble those of the 1D doped Hubbard model~\cite{aben06} and 
exhibit the features that are expected for electronic Luttinger liquids.
Spin and charge excitations are gapless with linear dispersions $\omega = v_{\text{c,s}} k_x$ at low energy.
The charge and spin velocities deduced from the CT-INT spectra are compatible with  
those obtained with the DMRG (see Fig.~\ref{fig:velocities}). 

\begin{figure}
\includegraphics[width=0.5\textwidth]{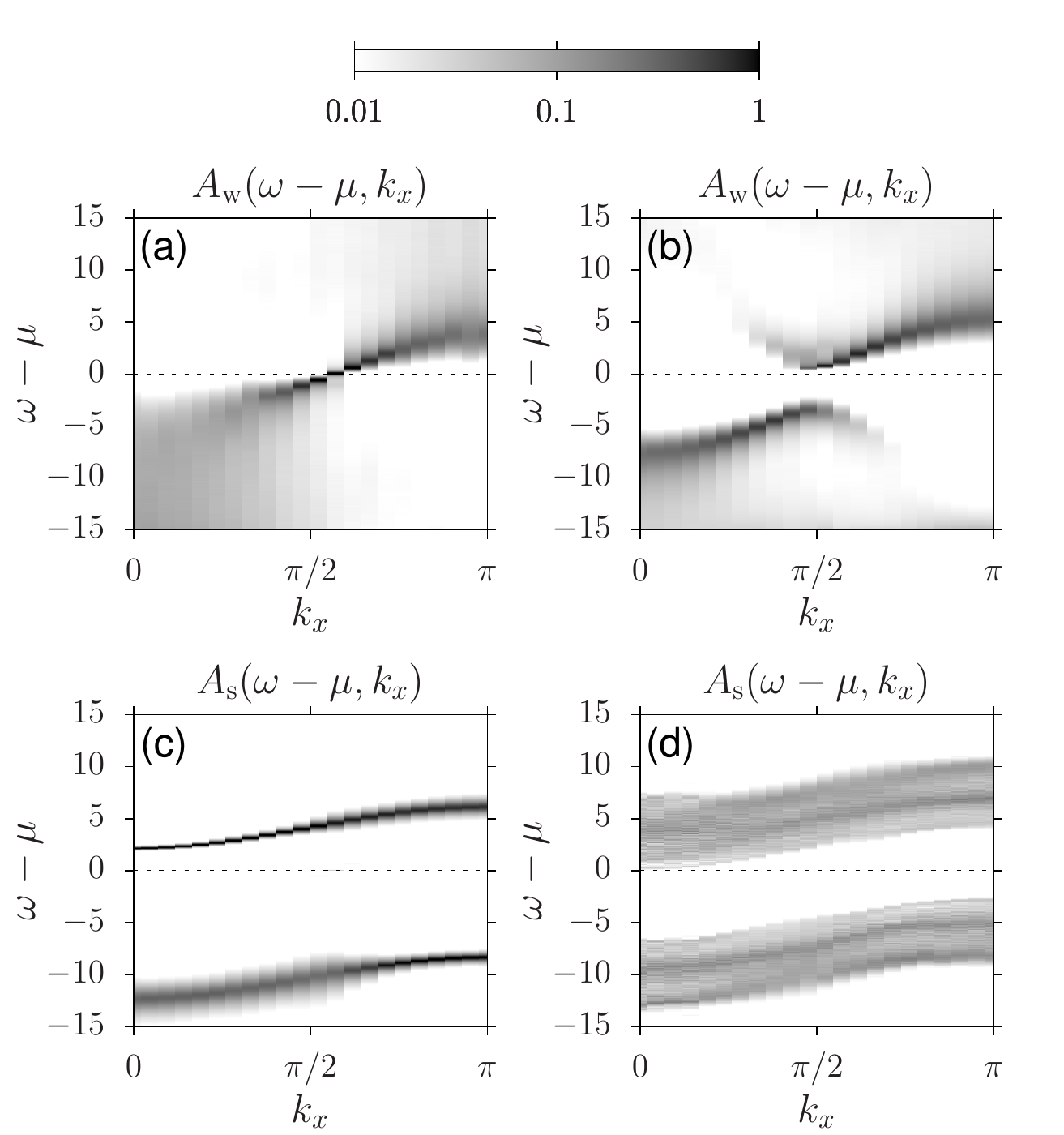}
\caption{\label{fig:qmc_U12_doped}
CT-INT results for the spectral functions $A_{\text{w}}(\omega,k_x)$ [(a),(b)] and
$A_{\text{s}}(\omega,k_x)$ [(c),(d)]  for
$U=12$, $\beta =10$, and $L_x=42$. Panels (a) and (c) show
results for the three-leg NLM, panels (b) and (d) for the 3D wire-substrate model ($L_y=42$, $L_z=10$). 
The chemical potential was $\mu=3.08$ for the NLM and $\mu=1.205$ for the 3D
model, corresponding to a doping of $y_\text{w}\approx 12.5\%$.
}
\end{figure}

The DMRG results for the three-leg NLM revealed a crossover between two conducting phases
upon increasing $U$ at fixed doping $y_{\text w}$. 
This crossover has not been investigated directly with the CT-INT method
because the critical coupling $U_c \approx 20$ is too large. On the other hand,
the critical coupling $U_c\approx 9$ in the 3D wire-substrate model is
small enough to carry out CT-INT simulations.
In that case, however, the chemical potential $\mu$ must be just above the lower edge of
the conduction band (or, equivalently, just below the upper edge of the valence band)
to achieve a finite wire doping $y_{\text w}$ but a vanishing dopant density in the wire
$y_{\text{eff}}$,  i.e., $\vert \mu\vert \agt \Delta_\text{s}/2$ for $U=0$. Finding the correct value of $\mu$ 
for $U>0$ turned out to be a rather delicate problem.

As an example, Fig.~\ref{fig:qmc_U12_doped} shows the single-particle spectral functions 
of the three-leg NLM and the 3D wire-substrate model away from half-filling for $U = 12$.
This interaction is below the critical value $U_c\approx 20$ of the three-leg NLM determined 
with DMRG but above the estimated critical value $U_c\approx 9$ for the 3D wire-substrate model.
(The other parameters are equal to those used in Fig.~\ref{fig:qmc_U12} for half-filling.)
Accordingly, we see that the spectral functions of the three-leg NLM are qualitatively similar to those
for $U=8$ in Fig.~\ref{fig:qmc_U8_doped}.
For the 3D wire-substrate model, however, Fig.~\ref{fig:qmc_U12_doped}(b) shows that the wire spectral function resembles that
for half-filling in Fig.~\ref{fig:qmc_U12}. The Fermi energy still lies within the Hubbard gap,
close to the bottom of the upper Hubbard band. In addition, Fig.~\ref{fig:qmc_U12_doped}(d) confirms that
the Fermi energy lies at the edge of the conduction band.
This corresponds to a doped band insulator with gapless single-particle
excitations delocalized
in the full substrate. The density of charge carriers is $y_{\text{w}}/N_{\text{leg}}$ in the NLM and thus vanishingly small 
in the 3D wire-substrate model.

\begin{figure}
\includegraphics[width=0.5\textwidth]{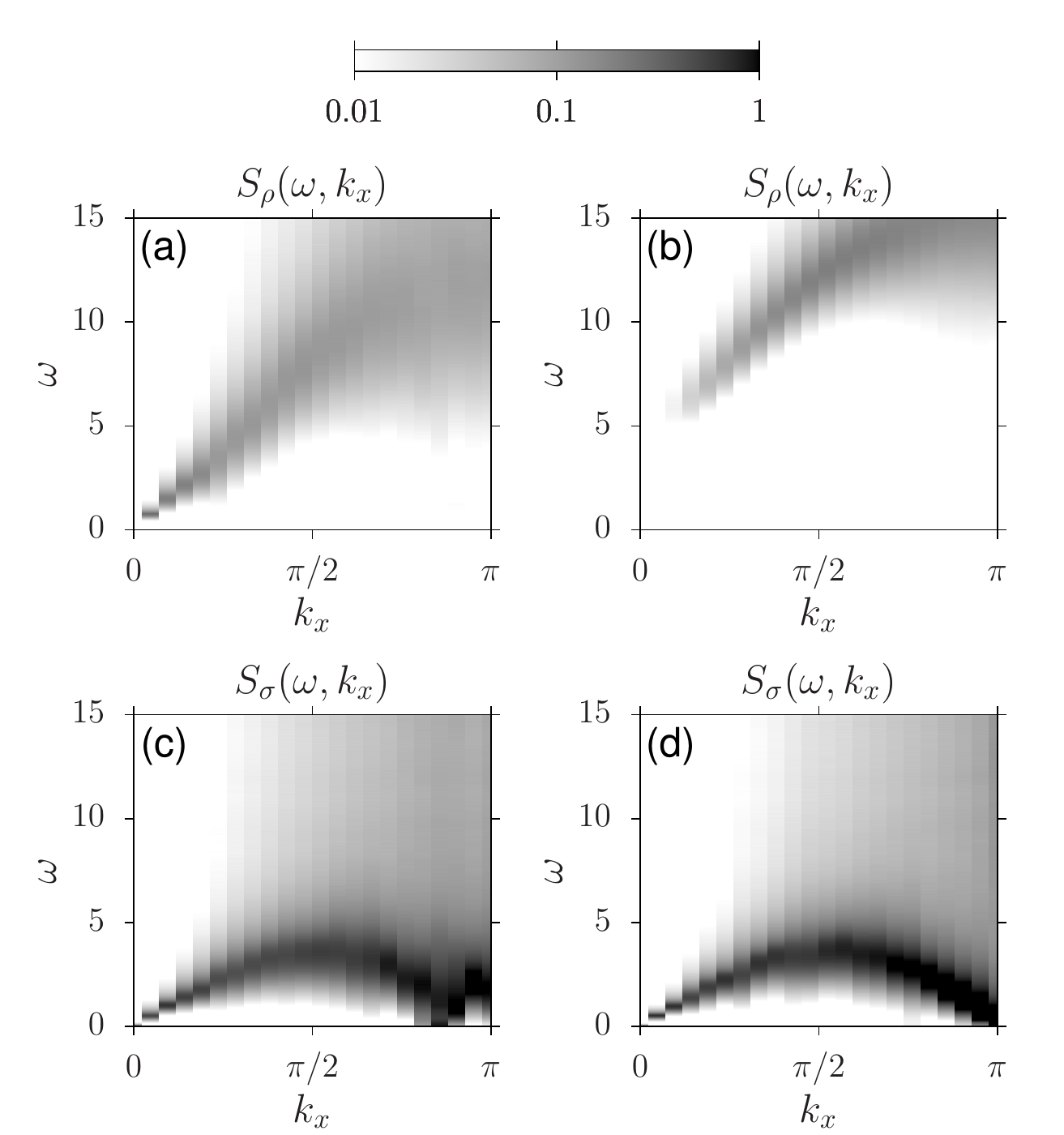}
\caption{\label{fig:qmc_U12_doped2}
CT-INT results for the dynamic charge structure factor $S_{\rho}(\omega,k_x)$
[(a),(b)] and the dynamic spin structure factor $S_{\sigma}(\omega,k_x)$ [(c),(d)]
on the wire for the same parameters as in Fig.~\ref{fig:qmc_U12_doped}.
Panels (a) and (c) show results for the three-leg NLM, panels (b) and (d) for the 3D wire-substrate model.}
\end{figure}

Figure~\ref{fig:qmc_U12_doped2}  shows the dynamic charge and spin structure factors of the wire 
for the same parameters as in Fig.~\ref{fig:qmc_U12_doped}.
Figures~\ref{fig:qmc_U12_doped2}(a) and~(b) confirm that charge excitations in the wire are gapless
for the three-leg NLM but have a gap equal to the Mott gap of the half-filled 3D wire-substrate model,
cf. Fig.~\ref{fig:qmc_U12b}(b). The spin excitations are gapless and the spin structure factors
are very similar in both models [Figs.~\ref{fig:qmc_U12_doped2}(c) and~(d)]. These results
confirm that the wire is a Luttinger liquid in the three-leg NLM but a half-filled Hubbard chain
in the 3D wire-substrate model in this particular parameter regime.

The differences between the three-leg NLM and the 3D wire-substrate model in Figs.~\ref{fig:qmc_U12_doped}
and~\ref{fig:qmc_U12_doped2}  can also be seen as an illustration
of the failure of the NLM approximation for metallic substrates found in~\cite{paper1}.  We see here that not only the substrate properties
but also the wire properties are not reproduced correctly by the NLM. 
Note, however, that the discrepancies are essentially due to 
the strong dependence of the effective substrate band gap $\Delta_{\text s}(N_{\text{leg}})$, and thus of the critical coupling $U_c$,
on the number of legs in the NLM.
So a possible remedy could be to rescale $\Delta_{\text s}(N_{\text{leg}})$ [i.e, to change the rung hoppings $t^{\text{rung}}_{n}$ for $n\geq 2$
in the Hamiltonian~(\ref{eq:ladder-hamiltonian})].

We conclude that in the doped three-leg NLM a transition occurs from a correlated quasi-1D gapless phase (Luttinger liquid) for $U \alt U_c$
to a doped band insulator for $U \agt U_c$. 
The DMRG results for $N_{\text{leg}}>3$ and the QMC spectra suggest that this transition is not an artifact of the NLM but
a feature of the 3D wire-substrate model that is qualitatively reproduced by the NLM.
This transition between 1D and 3D metallic phases 
is consistent with the transition from a quasi-1D Mott insulator to a 3D band insulator found at $U_c$ for half-filling.
 
Isolated correlated 1D conductors are Luttinger liquids~\cite{sol79,giamarchi07,Schoenhammer}.
As the NLM with a finite number of legs is a quasi-1D system, it is not
surprising that we find a gapless Luttinger liquid phase. However, 
it is far from obvious that another metallic phase would occur.
On the one hand, the existence of the Luttinger liquid phase for $U \alt U_c$ is fully supported by 
the confinement of low-energy excitations to the wire (Figs.~\ref{fig:densitiesAFHF} and~\ref{fig:qmc_U8_doped})
and their clear quasi-1D finite-size scaling (Fig.~\ref{fig:chrgspingaps}) with distinct charge and spin velocities 
(Figs.~\ref{fig:velocities} and~\ref{fig:qmc_U8_doped2}).
On the other hand, the existence of the uncorrelated metallic substrate phase for $U \agt U_c$ is inferred from the disappearance of these features,
in particular the delocalization
of excitations on the noninteracting substrate legs and the equality of finite-size charge, spin, and single-particle gaps.
While we could also interpret the metallic substrate phase of few-leg NLMs as
a Luttinger liquid with very weak effective interactions, such an
interpretation breaks down in the limit $N_{\text{leg}} \gg 1$ and hence in the 3D wire-substrate model.
 
Finally, we note that the differences between spin and charge velocities in the Luttinger liquid phase 
become smaller with increasing $t_{\text{ws}}$
(see Fig.~\ref{fig:velocities})
and thus the distinction between a weakly-coupled Luttinger liquid and a quasi-1D Fermi gas 
becomes moot in the limit $t_{\text{ws}} \rightarrow \infty$.
In contrast, any local measurement on the wire in the doped band insulator, such as the local DOS measured by  scanning tunneling spectroscopy,
could reveal a 1D Mott insulator, as correlation functions (see  Fig.~\ref{fig:correl}) suggest
that this state subsists as a correlated chain impurity embedded in the substrate for $U \agt U_c$.

\section{\label{sec:conclusion}Conclusions}

We investigated a correlated wire with a Hubbard interaction deposited on an insulating substrate
using the NLM approach developed in \cite{paper1}. Using the DMRG method, we
were able to determine the ground-state properties and gaps of NLMs with
different numbers of legs. The CT-INT QMC method was used to obtain the
spectral properties of both the three-leg NLM and the 3D wire-substrate
model. We found that a three-leg NLM already yields
a qualitative description of the low-energy physics of the full 3D wire-substrate system. 
A quantitative description (e.g., for the charge and spin velocities of the Luttinger liquids)
is possible when the low-energy excitations are localized on the wire and
DMRG calculations can be carried out for several numbers of legs. It would
certainly be useful to obtain additional information from field-theoretical methods.

We found that Mott-insulating and Luttinger liquid phases, which are possibly relevant for atomic wires on semiconducting substrates, 
can be observed in the 3D wire-substrate system and are well captured
by the NLM. 
Transitions from 1D low-energy excitations to low-energy excitations delocalized in the substrate
can also be observed in the NLM but, by nature, the results depend quantitatively on the number of legs.
While the spectral properties calculated with the CT-INT method confirm
that these transitions also occur in the 3D wire-substrate system, we have
not yet obtained accurate results for, e.g., the critical values of the
interaction $U$ and the hybridization $t_{\text{ws}}$.

It may be surprising at first to find transitions from 1D correlated phases (Mott insulator, Luttinger liquid) to 
uncorrelated phases (band insulator, metal) upon increasing the interaction
$U$ between electrons or decreasing the hybridization $t_{\text{ws}}$ between wire and substrate.
However, it should be realized that we consider only the low-energy excitations 
and that the latter are not always associated with the strongest coupling in a system.
This is easily seen in the limits $U \gg t_{\text{ws}}$ or $t_{\text{ws}} \gg U$.

In conclusion, the 3D wire-substrate model with a Hubbard-type wire and the corresponding effective narrow ladder models
provide us with a promising approach to investigate correlation effects in atomic wires on semiconducting substrates.
This approach can be easily generalized to extended Hubbard Hamiltonians and electron-phonon models.
The model properties can be determined using the CT-INT and DMRG methods and additional information
could be obtained using other methods for 1D strongly correlated systems.

\begin{acknowledgments}
This work was supported by the German Research Foundation (DFG) through SFB
1170 ToCoTronics and the Research Unit \textit{Metallic nanowires on the atomic scale: Electronic
and vibrational coupling in real world systems} (FOR1700, grant No.~JE~261/1-1).
Some of the DMRG calculations were carried out on the cluster system
at the Leibniz Universit\"at Hannover.
The authors gratefully acknowledge the computing time granted by the John
von Neumann Institute for Computing (NIC) and provided on the supercomputer
JURECA \cite{Juelich} at the J\"{u}lich Supercomputing Centre.
\end{acknowledgments}

  \end{document}